\documentclass{aa}

\usepackage{amsmath}
\usepackage[inline]{enumitem}
\usepackage[dvipsnames]{xcolor}
\usepackage[super]{nth}
\usepackage{mathtools}
\usepackage{hyperref}
\usepackage{makecell}
\usepackage{graphicx}
\usepackage{txfonts}
\usepackage{CJKutf8} 
\usepackage{stfloats}
\usepackage{subcaption} 
\usepackage{lscape} 
\usepackage{placeins} 
\usepackage[normalem]{ulem}

\newcommand{\msun}{\ensuremath{\mathrm{M}_\odot}}
\newcommand{\zsun}{\ensuremath{\mathrm{Z}_\odot}}

\newcommand{\hmr}{\ensuremath{r_\mathrm{h}}}
\newcommand{\mcore}{\ensuremath{m_\mathrm{c}}}
\newcommand{\rcore}{\ensuremath{r_\mathrm{c}}}

\newcommand{\rtide}{\ensuremath{r_\mathrm{tid}}}
\newcommand{\rlagr}{\ensuremath{r_\mathrm{Lagr}}}
\newcommand{\avmass}{\ensuremath{M_\mathrm{av}}}
\newcommand{\mcl}{\ensuremath{M_\mathrm{cl}}}
\newcommand{\kmps}{\ensuremath{\mathrm{km~s}^{-1}}}

\newcommand{\nbo}{\textsc{Nbody6++GPU}}

\newcommand{\nbody}{\textit{N}-body}
\newcommand{\mcluster}{\textsc{McLuster}}
\newcommand{\fopax}{\textsc{Fopax}}
\newcommand{\petar}{\textsc{PeTar}}

\newcommand{\tref}[1]{Table~\ref{#1}}

\begin{document}

\title{Direct $N$-body simulations of rotating and extremely massive Population III star clusters}
\titlerunning{Massive, rotating Population III star clusters}
\authorrunning{Wu et al.}

\author{
      Kai Wu\inst{1}
 \and Ataru Tanikawa\inst{2}
 \and Francesco Flammini Dotti\inst{3,4,5}
 \and Marcelo C. Vergara\inst{1}
 \and Boyuan Liu\inst{6}
 \and \\ Albrecht W. H. Kamlah\inst{1,7}
 \and Manuel Arca Sedda\inst{1,8,9}
 \and Nadine Neumayer\inst{7}
 \and Rainer Spurzem\inst{1,10,11}\thanks{E-mail: spurzem@ari.uni-heidelberg.de}
}
\institute{
        Astronomisches Rechen-Inst., Zentrum f\"ur Astronomie, Univ. of Heidelberg,
        M\"onchhofstr. 12-14, 69120, Heidelberg, Germany  
        \and Center for Information Science, Fukui Prefectural University, 4-1-1 Matsuoka Kenjojima, Eiheiji-cho, Fukui 910-1195, Japan  
    \and Department of Physics, New York University Abu Dhabi, PO Box 129188 Abu Dhabi, UAE
    \and Center for Astrophysics and Space Science (CASS), New York University Abu Dhabi, PO Box 129188, Abu Dhabi, UAE
    \and
    Dipartimento di Fisica, Sapienza, Universit\'a di Roma, P.le Aldo Moro, 5, 00185 - Rome, Italy
    \and Inst. f\"ur Theoretische Astrophysik, Zentrum f\"ur Astronomie, Univ. Heidelberg, Albert Ueberle Str. 2, 69120 Heidelberg, Germany
    \and Max-Planck-Institut f\"ur Astronomie, K\"onigstuhl 17, 69117 Heidelberg, Germany
    \and Gran Sasso Science Institute (GSSI), 67100 Viale Francesco Crispi 7, L’Aquila (Italy)
    \and INFN, Laboratori Nazionali del Gran Sasso, 67100 Assergi (Italy)
    \and National Astronomical Observatories, Chinese Academy of Sciences, 20A Datun Rd., Chaoyang District, 100101, Beijing, China
        \and Kavli Institute for Astronomy and Astrophysics, Peking University, Yiheyuan Lu 5, Haidian Qu, 100871, Beijing, China
}

\date{Received ---; accepted ---}

\abstract
{}
{We present eight direct $N$-body simulations with \textsc{Nbody6++GPU} of extremely massive, initially rotating Population~III star clusters with $1.01\times 10^5$ stars.
}
{Our models include primordial binaries, a continuous initial mass function, differential rotation, tidal mass loss, updated fitting formulae for extremely massive metal-poor Population~III stars, and general-relativistic merger recoil kicks. We assess their impact on cluster dynamics. 
}
{
All runs form black holes below, within, and above the pair-instability gap, with multi-generation growth.
Faster-rotating clusters core-collapse earlier; post-collapse clusters host a rotating, axisymmetric subsystem of intermediate-mass black holes (IMBHs) at the centre and an expanding halo of lower-mass objects. 
Pair-instability supernovae and compact-object formation at $\sim$2-$3$ Myr sharply reduce total mass and a large fraction of the cluster’s angular momentum.
All Population III clusters in our simulations have the gravothermal-gravogyro catastrophe phase.
}
{
We confirm two of the hypothesized formation channels of galactic nucleus seed black holes: gravitational runaway mergers of black holes and of Population III stars, which core-collapse into IMBHs thereafter. A higher initial star cluster bulk rotation correlates with earlier core collapse and, in the event counts reported here, with more coalescences and collisions, as well as lower retained (compact) binary abundances. 
Initial bulk rotation is a primary control parameter of cluster evolution: faster rotation accelerates early angular-momentum transport, gravothermal collapse, mass segregation, and amplifies post-collapse expansion, which also favours the formation of a compact central IMBH subsystem. 

}


\maketitle
\nolinenumbers

\section{Introduction}
        \label{Introduction}

        There are three main formation channels of seed black holes (BHs) for galactic nuclei that are typically considered today \citep[see the reviews by][and sources therein]{Rees1984,Greeneetal2020}. 
        \begin{enumerate*}[label=(\roman*)]
                \item Gravitational runaway mergers between stars and compact objects happen throughout cosmic time in dense star clusters. They can be separated into a `fast' and a `slow' regime following \citet{Greeneetal2020}. In the fast regime (a couple of million years from star cluster formation and natal gas expulsion), gravitational runaway mergers can happen during early star cluster evolution, when stars evolve and merge either through binary stellar evolution or dynamical collisions \citep{PortegiesZwartMcMillan2002,Sakuraietal2017,Gielesetal2018,Reinosoetal2018,Reinosoetal2021,Wangetal2022b,Vergara2023,Vergara2025, Rantala2025}. The slow regime (around 100 Myr to a gigayear from star cluster formation and natal gas expulsion) is populated by gravitational runaway mergers that occur between BHs. Gravitational runaway mergers of both kinds are postulated to produce intermediate-mass BHs (IMBHs) with masses of order $10^2-10^4$~\msun{}. Both the fast and the slow regimes have been confirmed extensively by simulations of dense star clusters using various methods and both mechanisms are instrumental to growing IMBHs \citep[e.g.][]{Gierszetal2015,Rodriguezetal2019,ArcaSedda2019,ArcaSeddaetal2020a,dragon2-1,DiCarloetal2020a,DiCarloetal2020b,DiCarloetal2021,Rizzutoetal2021a,Rizzutoetal2022,Levequeetal2022b,Maliszewskietal2022, Vergara2025a, Paiella2025}.
        \item {Population III (Pop III) stars above the pair-instability mass gap can collapse directly into BH seeds with masses of $\sim 10^2$~\msun{} \citep[e.g.][]{BrommLarson2004,Bromm2013,Woosley2017,Haemmerleetal2020, Mestichelli2024}. Within the Pop III cluster, stars can collide and form an even more massive star, collapsing into a heavier BH seed \citep[e.g.][]{Katzetal2015,Sakuraietal2017,Reinosoetal2018,Reinosoetal2021,Vergaraetal2021, Tanikawaetal2022b,Wangetal2022b}.}
        {\citet{Reinoso2025} shows that IMBHs with masses of up to $10^4$~\msun{} are possible to form through stellar mergers and gas accretion.} 
                \item {A pristine gas cloud can avoid fragmentation, collapsing directly into a massive star that eventually forms a heavy seed with masses of around $10^4$-$10^6$~\msun{} \citep[e.g.][]{BrommLoeb2003,Begelmanetal2006,Begelmanetal2008,Begelman2010, Chon2025, vanDokkum2025}. Extremely massive stars with masses of $10^{5}$-$10^{6}~\msun{}$ can explode by the general relativistic instability supernova mechanism \citep[e.g.][]{ShibataShapiro2002,Sakuraietal2015,Uchidaetal2017}.} 

        \end{enumerate*}

        As was noted above, metal-poor Population (Pop) III star clusters at high redshifts ($z \gtrsim 10$) with a top-heavy initial mass function (IMF; e.g. \citealt{ShardaKrumholz2022}) are prime candidates to produce BH seeds for galactic nuclei and nuclear star clusters \citep[NSCs][]{Neumayeretal2020,Greeneetal2020,Askaretal2021,Askaretal2022,Schleicheretal2022}. The IMBHs that form in these clusters can grow even more massive through tidal disruption events over long timescales \citep{stone2017formation,Sakuraietal2019,Wang2025}. The single- and multi-generation mergers from the gravitational runaway process of BHs and other compact objects originating from Pop III star clusters will be relevant gravitational wave (GW) detection events \citep[e.g.][]{Schneideretal2000,Schneideretal2002,Schneideretal2003,Kinugawaetal2014,Kinugawaetal2015,Kinugawaetal2021a,Kinugawaetal2021b,Hartwigetal2016,Belczynskietal2017,Ngetal2022,Mestichelli2024,Liu2024,liu2024merging,wu2025pair-instability,wang2025detection}. {\citet{liu2025on-the-formation} shows that binary BH mergers in Pop III star clusters can produce GWs, similar to those observed in GW231123. \citet{Plunkett2025} develops observational strategies to constrain the stellar demographics of Pop III stars with next-generation telescopes,} especially for the proposed third-generation ground-based GW detectors, the Cosmic Explorer \citep[CE][]{Reitzeetal2019,Evansetal2021}, Einstein Telescope \citep[ET][]{Punturoetal2010a,Punturoetal2010b,Sathyaprakashetal2012,Maggioreetal2020,LiuBromm2020cL,LiuBromm2021}, and for future space-borne detectors operating in the millihertz and decihertz frequency bands, such as the laser interferometer space antenna \citep[LISA][]{AmaroSeoaneetal2013, AmaroSeoaneetal2017, AmaroSeoaneetal2022}, \citep[TianQin][]{TianQin:2015yph,Liu:2020eko}, or the decihertz GW observatory \citep[DECIGO][]{DECIGO2011, DECIGO2021}, where IMBHs are expected to be bright GW sources \citep[e.g.][]{MillerColeman2002,AmaroSeoaneetal2007, AmaroSeoane2018, Janietal2020, ArcaSeddaetal2020b, ArcaSeddaetal2021b}. 
    
        Using detailed binary population synthesis (BPS), \citet{Tanikawaetal2021a} conducted studies on the merger rate density of Pop III binary BHs below, within, and above the pair-instability mass gap from isolated binary stellar evolution. They find that mergers between two low-mass BHs (in their models, low mass means $M<50$~\msun{}) independent of mass ratio and semi-major axis distributions of the primordial Pop III binaries could be identified from the BHBH mergers observed by the (Advanced) Laser Interferometer Gravitational-Wave Observatory \citep[(a)LIGO][]{Aasietal2015,Abbottetal2018d,Abbottetal2019b}, (Advanced) Virgo Interferometer \citep[(a)Virgo][]{Acerneseetal2015,Abbottetal2018d,Abbottetal2019b} and by extension also the Kamioka Gravitational Wave Detector \citep[KAGRA][]{Abbottetal2018d,Abbottetal2020e,Akutsuetal2019}, although the predicted present-day (10~Gyr) merger rates would be comparatively low ($\sim0.1~\mathrm{yr}^{-1}\mathrm{Gpc}^{-3}$) with regard to the merger rate density inferred by the GWTC-4 catalogue of $15.4^{+6.0}_{-4.5}$~$\mathrm{yr}^{-1}\mathrm{Gpc}^{-3}$ \citep{the-ligo-scientific-collaboration2025upper}. Similarly, mergers between a low-mass BH and a high-mass BH (in their models, high mass means $M>130$~\msun{} due to the pair instability mass gap) or mergers between two high-mass BHs will be detectable using the aforementioned currently available GW detectors according to \citet{Tanikawaetal2021a}. However, the authors caution that if two conditions hold simultaneously, namely a wide minimum semi-major axis in primordial Pop III binaries due stellar expansion in the protostellar phases and fast stellar rotation that causes excitation of (quasi-)chemically homogeneous evolution of these stars, during which Pop III stars stay more compact than their non-rotating counterparts due to mixed helium throughout the star \citep[e.g.][]{Maeder1987,YoonLanger2005,deMinketal2009}. {Rapidly rotating Pop III stars \citep{Nandal2024, Tsiatsiou2024} have been proposed as a possible source of the high nitrogen abundance observed in galaxies at high redshift \citep[e.g.][]{Bouwens2010,Tacchella2023,Marques-Chaves2024, Naidu2025}. Rotation in Pop III stars induces strong internal mixing, which enlarges the convective core and results in brighter stars, improving the chances of detecting these primordial objects in the early Universe with JWST \citep{Hassan2025}.}
    
    Direct detections of Pop III stars, their remnants, or their host clusters remain lacking because they lie at very high redshift. \citet{Schaueretal2022} reported a candidate extremely massive Pop III star at $z=6.2$, but this requires confirmation. \citet{Vanzellaetal2020} claimed a Pop III complex at $z=6.629$ with the MUSE Deep Lensed Field targeting the Hubble Frontier Field (HFF) galaxy cluster MACS J0416. Using JWST's NIRCam and lensing, \citet{Fujimoto2025} identified a strong Pop III cluster candidate at $z=6.5$. {Another candidate is the high-redshift galaxy RX J2129–z8He, which exhibits He~II emission at $1640 \mathring{\mathrm{A}}$ \citep{Wang2024}}. Despite JWST, unambiguous direct observations remain difficult \citep{Rydbergetal2013,Katzetal2022}. \citet{deSouzaetal2013} suggest that several hundred JWST supernova detections could constrain the Pop III IMF. Future wide-field surveys with Euclid \citep{Laureijsetal2011,Tanikawaetal2022a} and Roman will likely outperform JWST in constraining direct-collapse BHs above the pair-instability gap because they cover larger areas \citep{LazarBromm2022,Vikaeusetal2022}.

    Because direct observations of Pop~III stars and their environments are lacking, the parameter space for their formation is poorly constrained \citep[e.g.][]{Klessen2019, Klessen2023}. \citet{Fraseretal2017} inferred a Pop~III IMF from 29 extremely metal-poor Pop~II stars that are expected to form in environments enriched by Pop~III SNe, and found a Pop~I/II-like IMF, but the small sample prevents strong conclusions from being drawn. 
    Theory and simulations generally predict top-heavy Pop~III IMFs; such IMFs favour XRB populations with brighter and more spatially irregular X-ray emission \citep{Sartorio2023}. Hydrodynamic models are therefore widely used to probe plausible Pop~III IMFs and their environmental dependence \citep[e.g.][]{HiranoBromm2017,HiranoBromm2018,Susa2019,Chonetal2021,Shardaetal2021,Latifetal2022}. The binary fraction and initial binary statistics remain highly uncertain \citep{Stacyetal2013,Liuetal2021a} and, being environment-dependent, affect massive-star and compact-object demographics \citep{Hiranoetal2018,Sugimuraetal2020,Kroupa2025}.

        The observational and theoretical uncertainties are inherited by the initial conditions of Pop III star cluster simulations. One such parameter concerns the degree of initial Pop III star cluster rotation. A set of direct \nbody{} simulations of initially rotating Pop III star clusters were performed by \citet[$N=10^3-10^4$ and evolution only up to $2~\mathrm{Myr}$]{Vergaraetal2021}. They used Miyamoto-Nagai models with flattening and rotation and found that not only the collision rate increases with increasing bulk rotation, but additionally the number of escapers is reduced the larger the initial rotation is. The low particle numbers, the short simulation time, and other simplifying assumptions, such as neglecting stellar evolution, make it difficult to generalize these results. In general, for simulations in collisional dynamics of rotating star clusters, distribution functions from, for example, \citet{Goodman1983a}, \citet{LongarettiLagoute1996}, \citet{EinselSpurzem1999}, and \citet{VarriBertin2012} are typically used. \citet{Kamlahetal2022-rotation} recently ran simulations of Pop II star clusters with rotating King models from \citet[using the 2D Fokker-Planck code \fopax{}; \citealt{EinselSpurzem1999,Kimetal2002,Kimetal2004,Kimetal2008}]{EinselSpurzem1999}. Apart from the formation and dissolution of a rotating bar of BHs that is related to the gravothermal-gravogyro catastrophe \citep{Hachisu1979,Hachisu1982,AkiyamaSugimoto1989,EinselSpurzem1999,Hongetal2013,Kamlahetal2022-rotation}, they found a possible dependence of BHBH binary abundances on the initial star cluster bulk rotation; for a fast-rotating model (rotating King model $W_0=6.0,~\omega_0=1.2$), they found significantly more BHBH binaries during a phase of star cluster evolution when BHBH mergers in the aforementioned ‘slow’ regime are relevant \citep{Greeneetal2020}. Similarly, \citet{Webbetal2019} found that increased initial star cluster rotation precipitates the formation of circularized BHBH binaries. Since Pop III are postulated to harbour many more (IM)BHs than Pop II star clusters, relative to their size, we tested whether earlier findings apply to Pop III clusters.
    
        Furthermore, the lack of concrete observations of Pop III stars adds uncertainty to their stellar evolution. The most impactful parameter that differentiates stars of the same masses in Pop I and Pop III populations is the metallicity \citep[therefore, Pop III stars are also referred to as extreme metal poor stars e.g. in][]{Tanikawaetal2020}. The lower the metallicity, the weaker the radiation-driven winds that affect massive stars. Generally, this statement also holds for pulsation-driven winds \citep{Nakauchietal2020}. In the regime of $Z/\zsun{}<10^{-4}$ both winds become negligibly small. However, it has been suggested that Pop III stars form with very high rotation rates, when the magnetic fields are negligible, and thus there is negligible magnetic braking \citep{Stacyetal2011,Stacyetal2013b,HiranoBromm2018}. The winds due to the weakened stellar magnetic field of such stars may be very powerful \citep{Liuetal2021b,Jeena2023}. 

    The lower the metal in a Pop III star, the more drastic its internal evolutionary changes, and the lower its opacity \citep{Ekstroemetal2008}. First, the lower the metallicity of a star, the more compact it is, because the line-driven radiation pressure diminishes. Furthermore, most massive Pop I/II stars are characterized by a red supergiant (RSG) evolutionary phase with convective envelopes, while most Pop III stars end with a blue supergiant (BSG) phase and radiative envelopes \citep{Tanikawaetal2020}. These properties affect binary stellar evolution. While Pop I/Pop II stars tend to undergo unstable mass transfer and common envelope evolution (CEE) in the RSG phase, BSGs from Pop III stars undergo stable mass transfer, so that less mass is ejected from the binary system. As a result, BHBH binaries from Pop III stellar populations can be more massive than Pop I/II counterparts, even ignoring wind mass loss, making Pop III star clusters a very attractive target of GW event detection and progenitor studies \citep[e.g.][]{Inayoshietal2017,Kinugawaetal2021a,Kinugawaetal2021b,Tanikawaetal2021a,Tanikawaetal2021c,Tanikawaetal2021d,Tanikawaetal2022b,Liu2024}. {Dense stellar environments substantially increase the fraction of mergers involving BHs with primary masses exceeding the pair-instability mass limit \citep{Mestichelli2024, Mestichelli2025}. Furthermore, shorter initial orbital periods lead to higher merger rates \cite{Santoliquido2023}, which can boost the production of GW sources in the early Universe.}
    
     In this paper, we present eight direct \nbody{} simulations of extremely massive, rotating Pop~III clusters, run with and without self-consistent general relativistic (GR) merger recoil kicks. The main novelty is the inclusion of initial bulk rotation. We examine how initial bulk rotation, state-of-the-art Pop~III stellar-evolution models, GR merger recoil kicks (GRKs), primordial binaries, a continuous IMF, and tidal-field mass loss jointly shape cluster dynamics and IMBH formation.
    
        The paper is structured as follows. Section \ref{Section:Methods} summarizes the methodology, the implementation of the GRKs, and the Pop III stellar evolution fitting formulae. 
         Section \ref{Section:Initial conditions} outlines the initial conditions. Section \ref{Section:Results} presents the results. In Sect. \ref{Section:Summary and conclusions} we summarize and discuss the work.

\section{Methods}
        \label{Section:Methods}
    
We used \nbo{} \citep{Spurzem1999,NitadoriAarseth2012,Wangetal2015} for the dynamical evolution of the Pop III clusters. The initial rotating conditions followed \citet{Kamlahetal2022-rotation} and were generated with \mcluster{} \citep{Kuepperetal2011a,Levequeetal2022a} and \fopax{} \citep{EinselSpurzem1999}; details are given in Appendix~\ref{Section:McLuster,fopax}.

Population III stellar evolution was treated with the low-metallicity recipes of \citet{Tanikawaetal2020} at $Z/\zsun{}=10^{-8}$, without radiation-, pulsation- \citep[e.g.][]{Nakauchietal2020}, or rotation-driven \citep[e.g.][]{Liuetal2021b} wind mass loss; further details are in Appendix~\ref{Sect: SSE, BSE, and Pop III stellar evolution}. 
Fitting formulae for relativistic merger recoil kick velocity and remnant mass and spin followed \citet{Campanellietal2007,Loustoetal2012,JimenezFortezaetal2017}, as summarized in Appendix~\ref{Sect: General relativistic merger recoil kicks}.  We refer to BHs with masses above 100~\msun{} as IMBHs. 

\section{Initial conditions}
        \label{Section:Initial conditions}
        \begin{table}
        \centering
        \small
                \caption{Initial parameters that are identical across all eight initial models for the \nbo{} simulations. 
        }
                \label{Initial_conditions}
                \begin{tabular}{|c|c|}
                        \hline \textbf{Quantity} & \textbf{Value} \\
                        \hline Particle number & $1.01\times 10^5$ \\
                        \hline Cluster mass & $8.135\times 10^6$~\msun{} \\
                        \hline Cluster metallicity & $Z/\zsun{}=10^{-8}$ \\
                        \hline Binary fraction $f_{\text{b}}$ & $1.0\%$  \\
                        \hline Half mass radius \hmr{} & $1.00$~pc  \\
                        \hline Tidal radius \rtide{} & $264.50$~pc   \\
                        \hline Virial status & virial equilibrium   \\
                        \hline IMF & \makecell{flat ($\alpha$=1) \\ $8.0-300.0$~\msun{} \\ \citep{LazarBromm2022}} \\
                        \hline Density model & \makecell{King model \citep{King1962}\\ $W_0=6.0$} \\
                        \hline \makecell{Binary eccentricity \\ distribution $f(e)$} & \makecell{Thermal ($f(e)\propto e$)} \\ 
                        \hline \makecell{Binary semi-major axis \\ distribution $f(a)$} & \makecell{uniform in $\mathrm{log}(a)$ \\ between 10~au and 100~au } \\
                        \hline \makecell{Binary mass ratio \\ distribution $f(q)$} &  \makecell{uniform distribution of mass ratio \\ {0.1$<q<$1.0} 
                \\ 
                \citep{Kobulnickyetal2014}}\\
                        \hline
                \end{tabular}
        \end{table}     
        \begin{table}
        \centering
        \centering
                \caption{Model identifiers, rotation parameters, and whether {simulations enable GR recoil kicks at relativistic mergers}. 
        }
                \label{Model_IDs}
        \begin{tabular}{|c|c|c|c|c|}
            \hline \textbf{Model ID} & \textsc{K}0.0 & \textsc{K}0.6 & \textsc{K}1.2 & \textsc{K}1.8 \\
            \hline \textbf{GR kicks?} & yes & yes & yes & yes \\
            \hline $\omega_0$ & $0.0$ & $0.6$ & $1.2$ & $1.8$ \\
            \hline \textbf{Model ID} & \textsc{NoK}0.0 & \textsc{NoK}0.6 & \textsc{NoK}1.2 & \textsc{NoK}1.8 \\
            \hline \textbf{GR kicks?} & no & no & no & no \\
            \hline $\omega_0$ & $0.0$ & $0.6$ & $1.2$ & $1.8$ \\
            \hline
        \end{tabular}
        \end{table}     
        
        {The mass function of Pop III star clusters is highly uncertain. Formation of Pop III star clusters has been simulated in cosmological mini-halos \citep[cf. e.g.][]{Stacyetal2016}, and corresponding star-cluster simulations have covered masses from $10^3$ to $10^5$~\msun{} \citep{liu2025on-the-formation,wu2025pair-instability}. However, \citet{Kimmetal2016} propose the formation of more massive Pop III clusters without dark matter directly from cooling gas clouds, with masses of up to $5\cdot10^5\msun{}$, and \citet{Liuetal2021a} find cluster masses reaching this value. \cite{Stacyetal2016} also discuss a possible Pop III cluster with a mass as high as $10^7\msun{}$. In our simulations we explore the very high-mass end (over $10^5$ objects, with 1\% binaries and a top-heavy IMF; see \tref{Initial_conditions}) for two reasons: first, to extend previously published simulations to higher masses; and second, to probe very high initial densities that may promote rapid formation of massive stars and massive BHs \citep{Vergara2025a}.}

   {We started from an unsegregated, unfractal King model} \citep{King1966b,GoodwinWhitworth2004} with $W_0 = 6.0$, initially in virial equilibrium. The half-mass radius was set to \hmr{}=1~pc. As is outlined in Appendix~\ref{Section:McLuster,fopax}, the initial model from \mcluster{} was then redistributed with a rotating King model, which is more compact than its non-rotating counterparts \citep{EinselSpurzem1999}. Therefore, the internal structural parameters such as the \hmr{} and \rcore{} changed in this initialization step from their original \mcluster{} \nbody{} distribution (see Fig. \ref{Global_properties}).
    
        We used a flat IMF of between 8.0 and 300.0~\msun{} \citep{LazarBromm2022}, which gave an initial crossing time of 0.0074~Myr and half-mass relaxation time of 6.869~Myr. Primordial binaries were paired with uniform mass ratios ($0.1<q<1.0$)\footnote{{The \citet{Kobulnickyetal2014} distribution adopts a random pairing for stars with $m\leq5~\msun{}$, but due to our IMF set-up, all stars have $m>5~\msun{}$.}} following previous studies \citep{Kiminkietal2012,SanaEvans2011,Sanaetal2012a,Sanaetal2013a,Kobulnickyetal2014}; their semi-major axes were uniformly distributed between 10 and 100~au and their eccentricities were thermal. The cluster metallicity was $Z/\zsun{}=10^{-8}$, following \citet{Wangetal2022b} and matching the lowest currently available metallicity for the fitting formulae of \citet{Tanikawaetal2020}. 

    {We adopted a primordial binary fraction of $f_{\mathrm{b}}=1\%$. A higher $f_{\mathrm{b}}$ at $N\simeq10^5$ is computationally expensive. The Pop~III binary fraction remains highly uncertain and likely depends on formation environment and feedback; hydrodynamic simulations can yield high multiplicities (tens of percent; e.g. \citealt{Stacyetal2013}), while other physical effects may reduce fragmentation, and thus the multiplicity (e.g. magnetic field; \citealt{sharda2025population}; UV, \citealt{hosokawa2016formation}). We therefore treated $f_{\mathrm{b}}=1\%$ as a limiting assumption that isolates the role of rotation in a controlled pilot suite and provides a conservative lower bound on contributions from primordial binary evolution channels (Type-I and Type-III; Sect. \ref{Section:Stars and compact objects}). We discuss the expected impact of higher $f_{\mathrm{b}}$ on IMBH formation in Sect. \ref{Section:Summary and conclusions}.}

            We placed the clusters on a circular orbit at 13.3~kpc following \citet{Kamlahetal2022-rotation}; this reproduced the mass-loss evolution of the eccentric NGC3201 orbit \citep[8.60--29.25~kpc, $e=0.55$,][]{Helmietal2020a}, assuming a point-mass MW of $1.78\times 10^{11}$~\msun{} and $v_c = 240.0$~\kmps{} \citep{Helmietal2020a,BobylevBajkova2020}, yielding an initial tidal radius of 264.50~pc. This set-up followed \citet{Kamlahetal2022-preparing,Kamlahetal2022-rotation} to facilitate a direct comparison. 
    {Our models adopted a static, Milky-Way–like tidal field as a controlled test; we emphasize that realistic high-z environments can produce time-dependent tidal tensors and repeated shocks -- effects that cosmological studies show can materially increase mass loss and modify angular-momentum evolution \citep[e.g.][]{renaud2015a-flexible}. A complementary study embedding our clusters in cosmological tidal histories is warranted.}

    We stopped the simulations at 500~Myr because \begin{enumerate*}[label=(\roman*)] \item the gravothermal--gravogyro catastrophe has largely subsided and the cluster has undergone core-collapse; \item most ZAMS stars, particularly the most massive ones, have reached their final stages; and \item environmental effects (e.g. cluster mergers) make long-term isolated evolution of Pop III clusters unlikely \citep[see e.g.][]{ArcaSeddaGualandris2018}. \end{enumerate*} Thus runs much longer than $\sim$10~Gyr (e.g. \citealt{Wangetal2022b}) are not always required.

            Particles were removed once they reached twice the current tidal radius from the density centre; they were then counted as `escapers'. The tidal radius was recomputed from the current cluster mass.
        Apart from the common set-up above, we ran four initial rotation parameters, $\omega_0$, with and without GRKs, for eight simulations in total (\tref{Model_IDs}). 

        \section{Results}
        \label{Section:Results}

\subsection{Stars and compact objects}
\label{Section:Stars and compact objects}

\begin{figure*}
    \centering
        \includegraphics[width=0.9\textwidth]{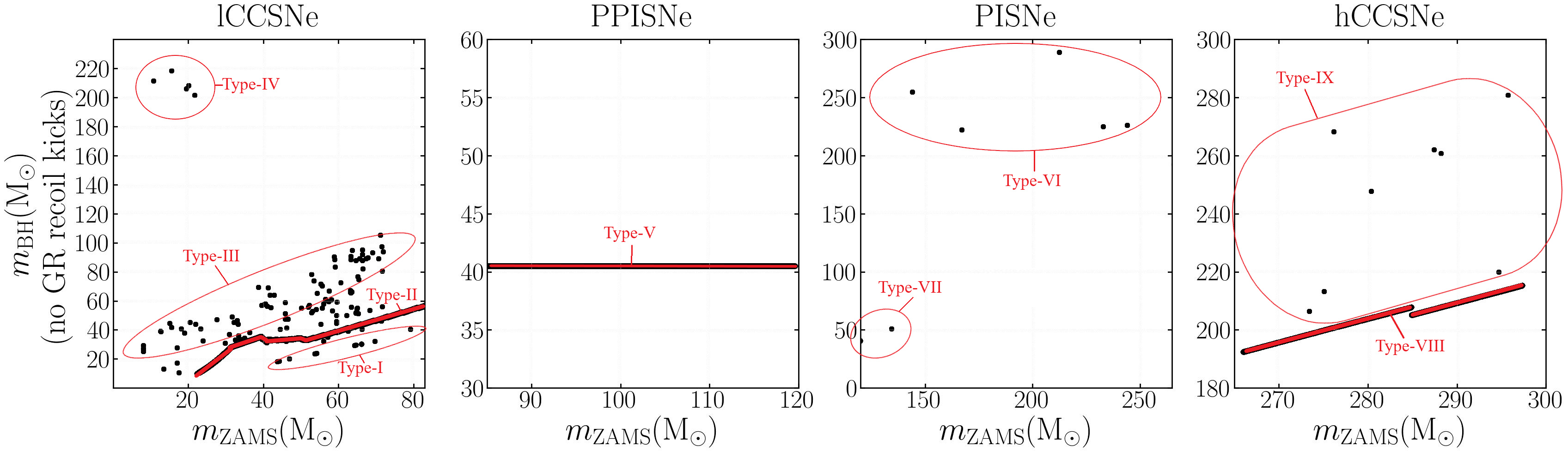}
        \caption{{Initial-final-mass relation (IFMR) for the \textsc{NoK}0.0 run}: BH mass, $m_{\mathrm{BH}}$, versus progenitor ZAMS mass, $m_{\mathrm{ZAMS}}$. The panels correspond to progenitors exploding as lCCSNe, PPISNe, PISNe, and hCCSNe. (IM)BH formation channels are shown in \textcolor{red}{red}.}
        \label{detailedIFMR}
\end{figure*}

Figure \ref{single_abundances_portrait} tracks counts of different stellar types.
Counts are similar during the first million years and then diverge as stellar evolution and core collapse proceed\footnote{Core collapse time is $\sim$50-80~Myr depending on $\omega_{0}$; see Sect. \ref{Section:Structural parameter evolution}.}. 
Larger $\omega_{0}$ speeds stellar loss, most clearly in \textsc{(No)K}1.8, while same-$\omega_{0}$ \textsc{K}/\textsc{NoK} pairs remain nearly identical. The trend reflects the gravothermal-gravogyro catastrophe combined with tidal mass loss.
By $\sim$20~Myr all MS stars have evolved and none remain. Near 20~Myr all models produce >1\,000 NSs ($n_{\mathrm{NS}}$), which later decline mainly because of escapers (Sect. \ref{Section:Escaper stars}). The BHs appear earlier ($\sim$6~Myr) and $n_{\mathrm{BH}}$ peaks at $\sim$50\,000 before 10~Myr.

We divided the ZAMS particles into four classes: low-mass core-collapse supernovae (lCCSNe):  $8.0 \leq m_{\mathrm{ZAMS}} < 85.0~\msun$, pulsational pair-instability supernovae (PPISNe):  $85.0 \leq m_{\mathrm{ZAMS}} < 120.0~\msun$, pair-instability supernovae (PISNe):  $120.0 \leq m_{\mathrm{ZAMS}} < 265.0~\msun$, and high-mass core-collapse supernovae (hCCSNe): $265.0 \leq m_{\mathrm{ZAMS}} < 300.0~\msun$, where $m_{\mathrm{ZAMS}}$ is the ZAMS stellar mass of a single star\footnote{A primordial binary may have members of two different mass groups.}. 
Figures \ref{detailedIFMR} and \ref{BH_IFMR} show the initial-final-mass relation (IFMR) for BHs retained in the cluster at 500~Myr, with $m_{\mathrm{BH}}$ (\msun{}) versus the progenitor ZAMS mass, $m_{\mathrm{ZAMS}}$ (\msun{}). Progenitors produce BHs via lCCSNe \citep{Fryeretal2012}, PPISNe \citep{Belczynskietal2016}, or hCCSNe \citep{Fryeretal2012}; PISNe leave no remnant \citep{Belczynskietal2016}\footnote{See also Appendix~\ref{Sect: SSE, BSE, and Pop III stellar evolution} and \citet{Kamlahetal2022-preparing}.}. 
Star-star and star-BH mergers produce IMBHs inside the gap; IMBHs are found above and below it.

We distinguish remnant-mass regimes of (IM)BHs produced by single versus binary evolution. Figure \ref{detailedIFMR} shows these regimes for the \textsc{NoK}0.0 simulation\footnote{Stellar type acronyms follow \citet{Hurleyetal2000}; also see Sect.~\ref{Sect: SSE, BSE, and Pop III stellar evolution}. }.
\begin{itemize}
    \item Type-I: The BHs from primordial binaries where H-envelopes of one or both members are fully stripped (by tides, RLOF, or CEE; see Appendix~\ref{Sect: SSE, BSE, and Pop III stellar evolution}), leaving naked HeMS cores that collapse into BHs. No mergers occur in this channel (c.f. Type-III); survivors may form a BH-BH binary depending on fallback and natal kicks.
    \item Type-II: Higher-mass BHs from single-star evolution using the delayed core-collapse SN model of \citet{Fryeretal2012}.
    \item Type-III: The BHs from primordial-binary mergers (CHeB+MS or CHeB+CHeB during RLOF); the merger product becomes a ShHeB star that later collapses to a BH.
    \item Type-IV: Dynamically formed IMBHs that always involve RLOF between an IMBH and a ShHeB star. 
    \item Type-V: The IMBHs from the PPISN channel (see Fig. \ref{BH_IFMR}; e.g. \textsc{K}0.0). 
    \item Type-VI: Two main dynamical channels for IMBH formation: (1) hyperbolic MS+MS collisions producing products that evolve to CHeB/ShHeB and then collapse to BHs; (2) hyperbolic coalescence of a CHeB+BH dynamical binary. 
    \item Type-VII: Hyperbolic CE events in dynamical CHeB/ShHeB binaries whose H-envelopes are stripped off and expelled. The cores merge and the product is a CHeB star following \citet{Hurleyetal2002b}. The product star then evolves into a ShHeB star and core-collapses into a BH. 
    \item Type-VIII: IMBHs form by single stellar evolution through core-collapse or the extension of the remnant mass functions of \citet{Fryeretal2012}.
    \item Type-IX: Dynamically MS/CHeB binaries undergo hyperbolic {collision} or coalescence following a CE event. Technically in \nbo{}, the product is labelled by the more massive member.
    \item Type-X: Rare MS+CHeB hyperbolic {collisions} with significant H-envelope loss, which yield CHeB, evolve into a ShHeB, and core-collapses into an IMBH.
\end{itemize}

We identified the formation channels present in the simulations using the Pop III fitting formulae of \citet{Tanikawaetal2020} and our initial conditions. Full \nbody{} dynamics (instead of population synthesis) is crucial: many (IM)BHs result from interactions in dynamically formed binaries.

Within $\leq$20~Myr and during the pre-core collapse\footnote{Core collapse is at 50--80~Myr; see Fig. \ref{rlagr}}, the Pop III clusters' stellar population is radically altered. Compared to similar Pop II clusters \citep{Kamlahetal2022-rotation}, this evolution is more extreme due to the extreme initial conditions. All Pop~III cluster models produce (IM)BHs below, within, and above the pair-instability gap.

\subsection{Compact binary fractions and general properties}
\label{Section:Compact binary fractions}
        \begin{figure}
        \centering
                \includegraphics[width=0.78\columnwidth]{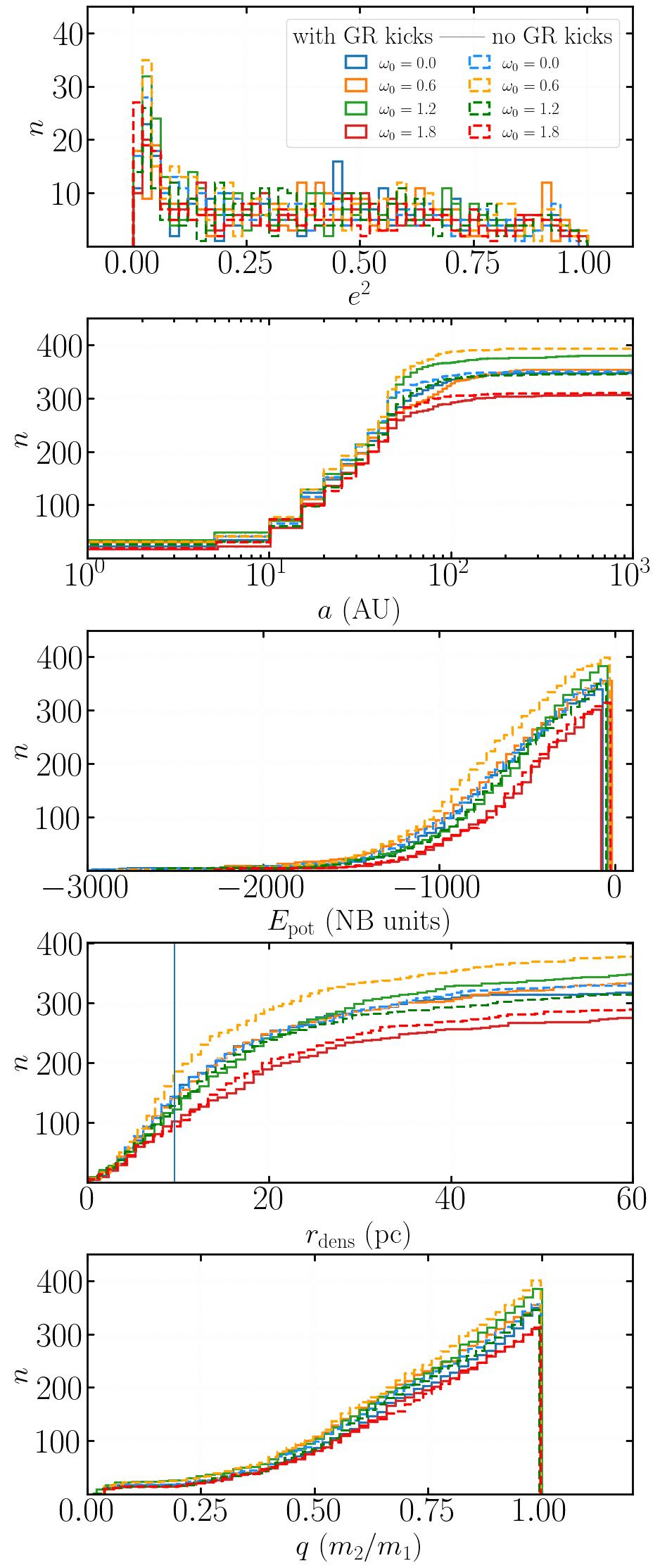}
                \caption{Binary distributions at 500~Myr. From top to bottom: $e^2$; cumulative semi-major axis, $a$ (au); cumulative binary potential energy, $E_{\mathrm{pot}}$ (\nbody{} units); cumulative distance to the density centre, $r_{\mathrm{dens}}$ (parsecs); and mass ratio $q\equiv m_2/m_1$. 
                In the $r_{\mathrm{dens}}$ panel a \textcolor{NavyBlue}{blue} line shows the half-mass radius \hmr{} of the \textsc{NoK}0.0 model at 500~Myr ($\sim$9.77~pc).}
                \label{binary_distributions}
        \end{figure}
    
The temporal evolution of compact-binary abundances (Fig. \ref{binary_abundances_portrait}; \tref{tab:binary_abundances}) traces the number of dynamical interactions. The \textsc{(No)K}1.8 models have the lowest binary abundances. The \textsc{(No)K}1.2 models show a comparable tendency, but their abundances lie close to those of the \textsc{(No)K}0.0 and \textsc{(No)K}0.6 models, so the trend is less clear. Notably, most abundances in the \textsc{(No)K}0.6 models exceed those in the \textsc{(No)K}0.0 models. See Sect. \ref{Section:Escaper stars} for a detailed analysis based on escaper statistics.

Figure \ref{binary_distributions} summarizes binaries within the tidal radius at 500~Myr. The $e^2$ distributions are bottom-heavy and similar across simulations. The cumulative $a$ distributions diverge around $a\approx50$~au: most binaries are within $a\approx50$~au in \textsc{NoK}1.8 but not in \textsc{NoK}0.6 or \textsc{NoK}1.2, indicating stronger disruption of wide binaries in the fastest-rotating model mostly before core-collapse (see Fig. \ref{binary_abundances_portrait}). 

The $E_{\mathrm{pot}}$ distributions show fewer hard binaries in \textsc{NoK}1.8. This can be explained by (i) prior collisions/coalescences (Sect. \ref{Collision and coalescence events}, Figs. \ref{binary_abundances_portrait} and \ref{time_evolution_ncoal_coll}, \tref{tab:binary_abundances}) and/or (ii) preferential escape of hard binaries in rapidly rotating runs (Fig. \ref{escapers_binaries}). This effect ejects many would-be merging binaries early, lowering in situ merger rates of stars and BHs. 

The \nth{4} panel of Fig. \ref{binary_distributions} shows the radial distribution of binaries at 500~Myr ($r_{\mathrm{dens}}$); the vertical line marks \hmr{} of the \textsc{NoK}0.0 model. Binary heating dominates post core-collapse evolution and produces the self-similar \rlagr{} evolution seen in Fig. \ref{rlagr}. Although higher initial rotation causes earlier core collapse via the gravothermal-gravogyro instability, by 500~Myr the binary radial distributions, if normalized to the retained binary counts, are similar across models, especially within \hmr{}. At 500~Myr, we therefore find no excess of harder binaries near the centres of faster-rotating clusters; long-term evolution largely erases the early structural (Figs.~\ref{Global_properties} and \ref{rlagr}) and angular-momentum differences, and equalizes the distributions of binaries that remain bound at 500~Myr.

The bottom panel of Fig. \ref{binary_distributions} shows the mass-ratio distribution $q\equiv m_2/m_1$ ($m_2<m_1$). Its shape is insensitive to initial rotation and to whether GR recoil kicks are included.

In the \textsc{(No)K}0.6 runs GRK reduces binaries by $\sim$50 (mainly at $a=40-50$~au, $r_\mathrm{dens}=10$~pc, $E_\mathrm{pot}=-1000$~\nbody{} units), while in \textsc{(No)K}1.2 it increases binaries by $\sim$30 (mainly at $a=50-60$~au, $r_\mathrm{dens}=30$~pc). The effects of GRKs in the non-rotating and fastest-rotating models are less visible. Overall, our results do not reveal a clear or systematic effect of GRKs on the binary distribution within the simulated clusters.

Note that the results are affected by small-number statistics, because each simulation contains only 1\,000 primordial binaries. More simulations are needed to reach robust conclusions.

        \subsection{Global dynamical evolution}
        \label{Section:Global dynamical evolution}
        \subsubsection{Structural parameter evolution}
        \label{Section:Structural parameter evolution}
\begin{figure}
    \centering
        \includegraphics[width=0.8\columnwidth]{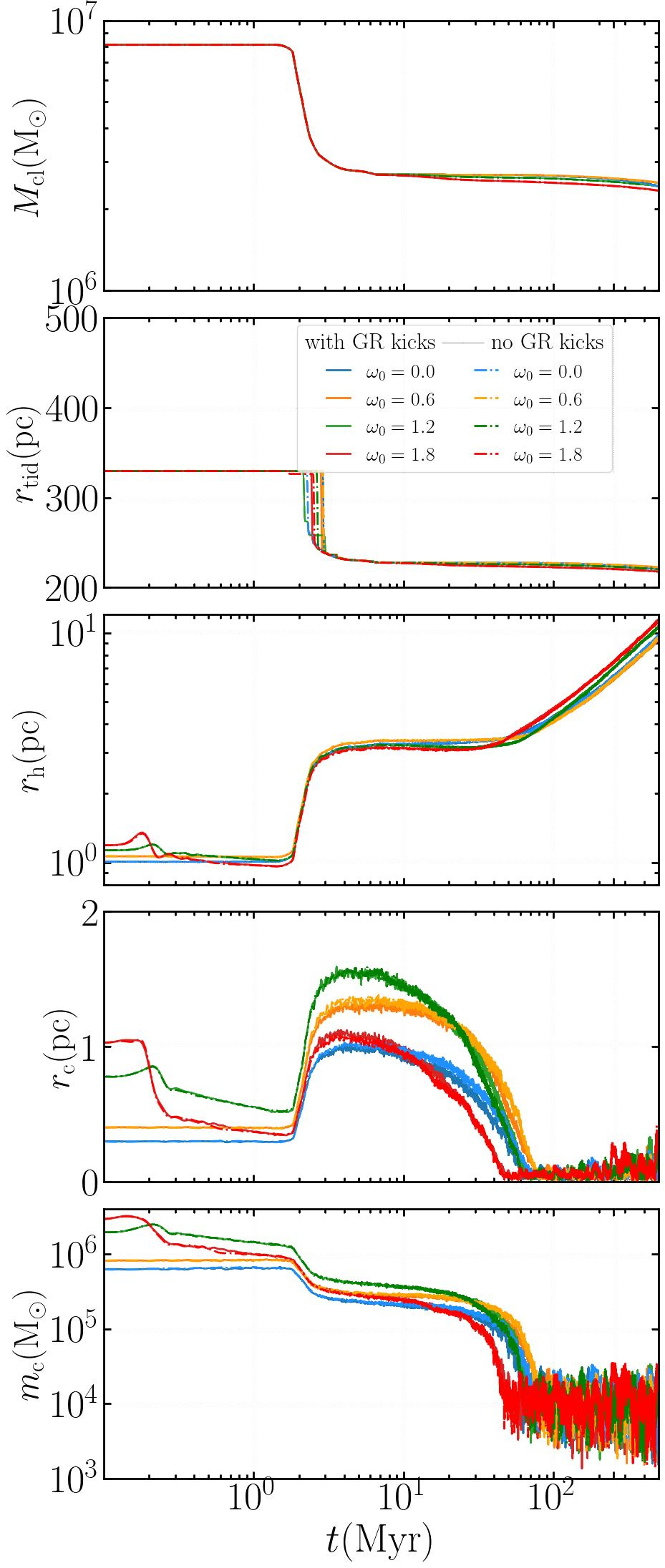}
        \caption{Total cluster mass, $\mcl{}$, tidal radius, \rtide{}, half mass radius, \hmr{}, mass of the core, $\mcore{}$, and radius of the core, \rcore{}, in the four panels for all eight simulations with and without GR recoil kicks for different $\omega_{0}$. Time is shown on a logarithmic scale to highlight the cluster's rapid early evolution. The models with \textsc{K} are plotted as solid curves and the models without (\textsc{NoK} models) are plotted as dash-dotted curves.
    }
        \label{Global_properties}
\end{figure}

We ran each of the four initial models ($\omega_{0} = 0.0, 0.6, 1.2, 1.8)$ using \nbo{} with GRKs switched on (\textsc{K} models) and without (\textsc{NoK} models). Models with the same $\omega_{0}$ evolve similarly whether or not GRKs are enabled, so the dominant control parameter of the global evolution is the initial bulk rotation. Figures \ref{Global_properties} and \ref{rlagr} show a four-phase evolution:

\begin{itemize}
    \item Phase I: Before the stellar evolution mass loss begins to dominate the star cluster evolution, namely, $0 - 2$~Myr. \mcl{} and \rtide{} stay constant, because mass loss through escapers or stellar evolution is tiny. Depending on $\omega_0$, the other structural parameters in Fig. \ref{Global_properties} show many more variations. \citet{Kamlahetal2022-rotation} used the same rotating King model set-ups for Pop II clusters, and showed that angular-momentum transport is very rapid and usually occurs before core collapse; the larger $\omega_0$, the faster this phenomenon happens. In the Pop III clusters in this work, we also observe this fast angular momentum transport, 
    {for example in the panel showing the half-mass radius} of Fig. \ref{Global_properties}. Within the first million years, the larger $\omega_0$, the more pronounced and earlier is the temporary increase in \hmr{} and decrease shortly after. This evolution is even more apparent in Fig. \ref{rlagr}, which shows the evolution of the Lagrangian radii \rlagr{} within spheres that contain 1\%, 5\%, 10\%, 30\%, 50\%, and 90\% of the total cluster mass at the current simulation time step. For example, in the \textsc{(No)K}1.8 models
    , the degree of initial bulk rotation has a pronounced effect: the outermost \rlagr{} expands substantially, while the other Lagrangian radii generally contract up to 2~Myr. All \rlagr{} show a small and synchronous bump (i.e. a brief increase followed by a decrease) at around 0.1~Myr. The \textsc{(No)K}1.2 models show the same effect, but weaker and slightly delayed relative to the \textsc{(No)K}1.8 models. Such a phenomenon is not shown in \textsc{(No)K}0.0 and \textsc{(No)K}0.6 models, where \hmr{} stays roughly constant until 2~Myr. This early evolution is also mirrored by \rcore{} and \mcore{}, where both the \textsc{(No)K}1.2 and the \textsc{(No)K}1.8 exhibit fast increases with subsequent decreases in the respective models. Again, this evolution is more pronounced and happens earlier in the \textsc{(No)K}1.8 with respect to the \textsc{(No)K}1.2 models. In summary, from $0$ to $2$~Myr, the evolution is governed only by the initial bulk rotation; tidal and stellar-evolution mass losses are negligible. 

    \item Phase II: Rapid stellar evolution mass loss around $2-3$~My r, as shown in \nth{1} panel of Fig. \ref{Global_properties}. This is mainly due to compact object formation, particularly PISN events, which produce no BHs and lose all the mass. Besides stellar evolution, mass loss is also caused by (a) escaping stars, through compact object natal kicks or GRKs, or (b) dynamical encounters, including strong encounters or by cumulative weak encounters. The cluster kinematics also change here owing to fallback-dependent natal kicks, which are generally larger for lower-mass BHs \citep[see][]{Kamlahetal2022-preparing}. As a result of the total mass drop, \rtide{} drops, and \hmr{} with \rcore{} rises sharply in all models. Interestingly, the maxima of \rcore{} lie in the \textsc{(No)K}1.2 model, while the cores of the \textsc{(No)K}1.8 and \textsc{(No)K}0.0 models are similarly compact. 
    We compare this with the average mass in the core \avmass{} within different Lagrangian radii
     \rlagr{}, shown in Fig. \ref{rlagr}. Signified by \avmass{}, we see that here all models show a similar segregation of masses into different spheres independent of initial bulk rotation; for example, the minimum \avmass{} in the 10\% \rlagr{} has similar values occurring at similar times, although we note that this quantity is subject to large statistical fluctuations.
    
    \item Phase III: From the end of phase II ($\sim$3~Myr) to the time of gravothermal core-collapse. As shown in Fig. \ref{rlagr}, the ending time of this phase varies between models, which is $\sim$80~Myr for \textsc{(No)K}0.0, $\sim$70~Myr for \textsc{(No)K}0.6, $\sim$60~Myr for \textsc{(No)K}1.2, and $\sim$50~Myr for \textsc{(No)K}1.8. Therefore, the larger the degree of initial bulk rotation in our Pop III clusters, the higher the central density, and the shorter the gravothermal collapse timescale, as seen in previous studies \citep[e.g.][]{AkiyamaSugimoto1989,EinselSpurzem1999,Kimetal2002,Kimetal2004,Kimetal2008,Fiestasetal2012,Kamlahetal2022-rotation}. Larger $\omega_0$ leads to higher \avmass{} in the 5\% and 10\% spheres at core collapse. Thus, stronger initial rotation accelerates mass segregation. In \textsc{(No)K}1.8 models, the cluster expansion ejects many stars\footnote{See the next paper in this series.}, producing a decrease in the 90\% \rlagr{}. 

    \item Phase IV: From core-collapse to 500~Myr (the simulations' end). The Lagrangian radii evolve self-similarly, as shown in \ref{rlagr}, driven by binary heating in the core which causes global expansion. Models with larger $\omega_0$ expand more by 500~Myr because they undergo earlier core collapse; their subsequent \rlagr{} evolution is otherwise similar. The clusters become fully mass-segregated (\avmass{} $\approx 200~\msun{}$ centrally, $\approx30~\msun{}$ in the halo). The same long-term trends appear in Fig. \ref{Global_properties}; longer runs may amplify differences. These results are consistent with energy-flow theorems across the half-mass radius \hmr{} \citep{Henon1975,BreenHeggie2013}.
\end{itemize}

The gravothermal-gravogyro catastrophe is therefore present in all our Pop~III cluster models, but its timescale depends strongly on the initial rotation. The GRKs have little effect on the global structural evolution.

\subsubsection{Angular momentum transport}
\label{Section:Angular momentum transport}

We explored whether and how angular-momentum redistribution depends on stellar evolution, GRKs, and initial bulk rotation. 
We tracked particles from each group over the full evolution and computed their total angular momentum squared $L_{\mathrm{group}}^2$ as \citet{Kamlahetal2022-rotation} normalized by $L^2_{\omega_{0}=0.6,t=0}$ for the \textsc{(No)K}0.6 model. Figure \ref{avmass_l^2} shows these quantities together with the group masses, $M_{\mathrm{group}}$.

Phase I: $0$ to $\sim$2~Myr, before stellar-evolution mass loss dominates. $M_{\mathrm{group}}$ evolves similarly across groups, but $L_{\mathrm{group}}^2/L^2_{\omega_{0}=0.6,t=0}$ does not. Higher $\omega_0$, notably \textsc{(No)K}1.8, clearly shows larger $L_{\mathrm{group}}^2$ for all mass groups, reflecting rapid global expansion and outward migration of stars. This matches the outer-radius growth and inner-radius contraction visible in Figs.~\ref{Global_properties} and \ref{rlagr} (Sect. \ref{Section:Structural parameter evolution}).

Phase II: $\sim$2-3~Myr. Rapid stellar evolution drives a sharp fall in $M_{\mathrm{group}}$, with higher-mass groups declining earlier owing to their shorter evolution timescales. The PISNe remove both mass and a large fraction of angular momentum: $M_{\mathrm{PISNe}}$ carried the most angular momentum before these events and is effectively depleted afterwards, causing a large loss of rotational kinetic energy.

Phase III: $\sim$3~Myr to core collapse (around $\sim$50-80~Myr depending on $\omega_{0}$. See Sect. \ref{Section:Structural parameter evolution} and Fig. \ref{rlagr}). The quantity $L_{\mathrm{hCCSNe}}^2/L^2_{\omega_{0}=0.6,t=0}$ initially falls after phase~II but, in the \textsc{(No)K}1.2 and \textsc{(No)K}1.8 models, it instead rises sharply and peaks at $\sim$10-20~Myr. After the peak, the quantity of \textsc{(No)K}1.2 shows a gradual decline, while that of \textsc{(No)K}1.8 drops rapidly. Other mass groups follow the same pattern, except for $M_{\mathrm{PISNe}}$. In \textsc{(No)K}1.8 the quantity attains a local minimum at core collapse, when the cluster is essentially fully mass-segregated (Fig. \ref{rlagr}).

Phase IV: Core collapse to simulation end. In all models $L_{\mathrm{hCCSNe}}^2/L^2_{\omega_{0}=0.6,t=0}$ decreases while $L_{\mathrm{lCCSNe}}^2/L^2_{\omega_{0}=0.6,t=0}$ increases, indicating angular momentum transfer from high-mass to low-mass BHs. Low-mass BHs move to wider orbits and a compact IMBH subsystem forms at the centre (also see Fig. \ref{rlagr}). Combined with Sect. \ref{Section:Structural parameter evolution} and Fig. \ref{detailedIFMR}, this points to binary heating of the central IMBHs as the dominant driver of the cluster evolution in this phase.

Compared to rotating Pop~II clusters \citep{Kamlahetal2022-rotation}, the detailed transport history differs because of the IMF and stellar evolution, but the gravothermal-gravogyro catastrophe remains visible in both cases. The GRKs again produce no strong effect.

\subsection{Collision and coalescence events}
\label{Collision and coalescence events}

We defined a `coalescence' as the merger of a {nearly} circular binary, {while `collisions' are a direct physical impact either on a hyperbolic or highly eccentric orbit; this is also how the \nbo{} code separates them.}
Under our initial conditions and Pop III stellar evolution prescriptions, we {find} collisions of BHBHs, MSBHs, and MSMSs, while coalescences may involve MS, HG, CHeB, ShHeB, HeMS, NS, and BH progenitors. 
Figure \ref{MS_GB_CHeB_ShHeB_NS_BH_mcoalescence_RI} shows the product masses, $m_{\mathrm{coal}}$ and $m_{\mathrm{coll}}$, versus $r_{\mathrm{dens}}$. For coalescence, only MS, CHeB, ShHeB, and BH products appear, consistent with the merger-product treatment of \citet{Hurleyetal2002b} and \citet{Kamlahetal2022-preparing}. Event totals are given in \tref{tab:coalescence_collision_rates}.

{Our suite contains one realization per set-up ($N=1$ per model ID). For integer event counts, $N$ (specific coalescence and collision types, compact-object merger channels), the counting uncertainty is well approximated by Poisson statistics, with a reference $1\sigma$ uncertainty of $\sigma_N\simeq\sqrt{N}$ (relative uncertainty $\sigma_N/N\simeq 1/\sqrt{N}$). 
Consequently, differences of a few events are not, by themselves, strong evidence of a systematic dependence on $\omega_0$ when $N$ is in the single-digit to few-tens regime\footnote{By contrast, trends inferred from continuous structural quantities (e.g. Figs. \ref{Global_properties} and \ref{rlagr}, and the angular-momentum transport in Fig. \ref{avmass_l^2}) and from large-number counts (e.g. total escaper numbers; \tref{tab:escaper abundances}) are dynamically more robust in our dataset.}. }

The coalescence events show no marked spatial differences: most occur within the central 2.5~pc, i.e. inside $r_{h}$ during the first $\sim$40~Myr (Fig. \ref{Global_properties}). They are mostly completed by 4-5~Myr (pre-core-collapse; Fig. \ref{time_evolution_ncoal_coll}). Many of these early mergers populate the PISN mass gap, indicating that stellar mergers in early star cluster evolution can seed BHs in and around the gap \citep[see e.g.][for recent hydrodynamical simulations of stellar collisions on this topic]{Costaetal2022,Balloneetal2022}. With our flat IMF, Pop III stellar evolution, and absent wind mass loss, producing BHs above the PISN gap is easier here than in typical young massive clusters \citep[e.g.][]{Rizzutoetal2022}.

Most coalescences produce CHeB stars and the most massive products, typically occurring within a few million years in all runs. Faster initial rotation {is consistent with more CHeB coalescences (Fig. \ref{time_evolution_ncoal_coll}), but Poisson uncertainties and covariances between channels in a single realization make this indicative.} The numbers of BH, ShHeB, and MS coalescences are similar across simulations. These mergers produce stars (mainly CHeB) with masses inside and above the pair-instability gap; some IMBHs end up in the pair-instability gap after accreting stellar material and grow further.

Collisions of MSMS, BHMS and BHBH offer useful diagnostics. 
Collision-produced IMBH masses are broadly similar across models (Fig. \ref{MS_GB_CHeB_ShHeB_NS_BH_mcoalescence_RI}), except for \textsc{NoK}1.2 (IMBH masses $\approx$ 1000~\msun{}), which is not seen in \textsc{K}1.2. Figure \ref{time_evolution_ncoal_coll} suggests more MSMS and BHBH collisions at a higher initial rotation, but {given the small absolute counts the implied trend should be read as suggestive and may be affected by stochastic fluctuations.} Notably, \textsc{K} models show multi-generation IMBH growth.
    
Figure \ref{BHBH_MSMS_MSBH_collisions} shows the primary versus secondary masses for MSMS, BHMS, and BHBH collisions across all runs; the distributions are similar regardless of rotation or GRKs. Contrary to theoretical expectations \citep[e.g.][]{Morawskietal2018,Morawskietal2019}\footnote{From these theories, we would expect that mergers have a very low GRK velocity, which would make them promising candidates for growing seed BHs for galactic nuclei.}, Fig. \ref{q_vgw} demonstrates that most GRK velocities, $v_{\mathrm{GRk}}$, exceed cluster escape speeds in the core, so merger remnants are often ejected even for $q\approx1$. Hence, {IMBHs of $>1000\msun{}$ are unlikely to form via hierarchical merging in Pop III clusters unless additional processes retain the remnants\footnote{A surrounding dark matter halo is unlikely to help to retain merger remnants. Even assuming a NFW big-halo of 100 times of the cluster total mass with 10 times of the cluster tidal radius will only contribute to a few \kmps{} of the escape speed in the core.}, such as} strong gas inflow after cluster formation causing extreme contraction of the cluster \citep{Kroupaetal2020}. 

To summarize, with increasing initial rotation, $\omega_{0}$, coalescences occur earlier and more frequently, and there is a tentative rise in MSMS and BHBH collision counts limited by low-number statistics. Enabling GRKs generates kick velocities that frequently exceed the cluster escape speed, favouring remnant ejection and thus suppressing collision-built IMBHs, even though multi-generation IMBH growth can still occur.

\subsection{Escaper stars}
\label{Section:Escaper stars}

\begin{figure}
    \centering
    \includegraphics[width=0.9\columnwidth]{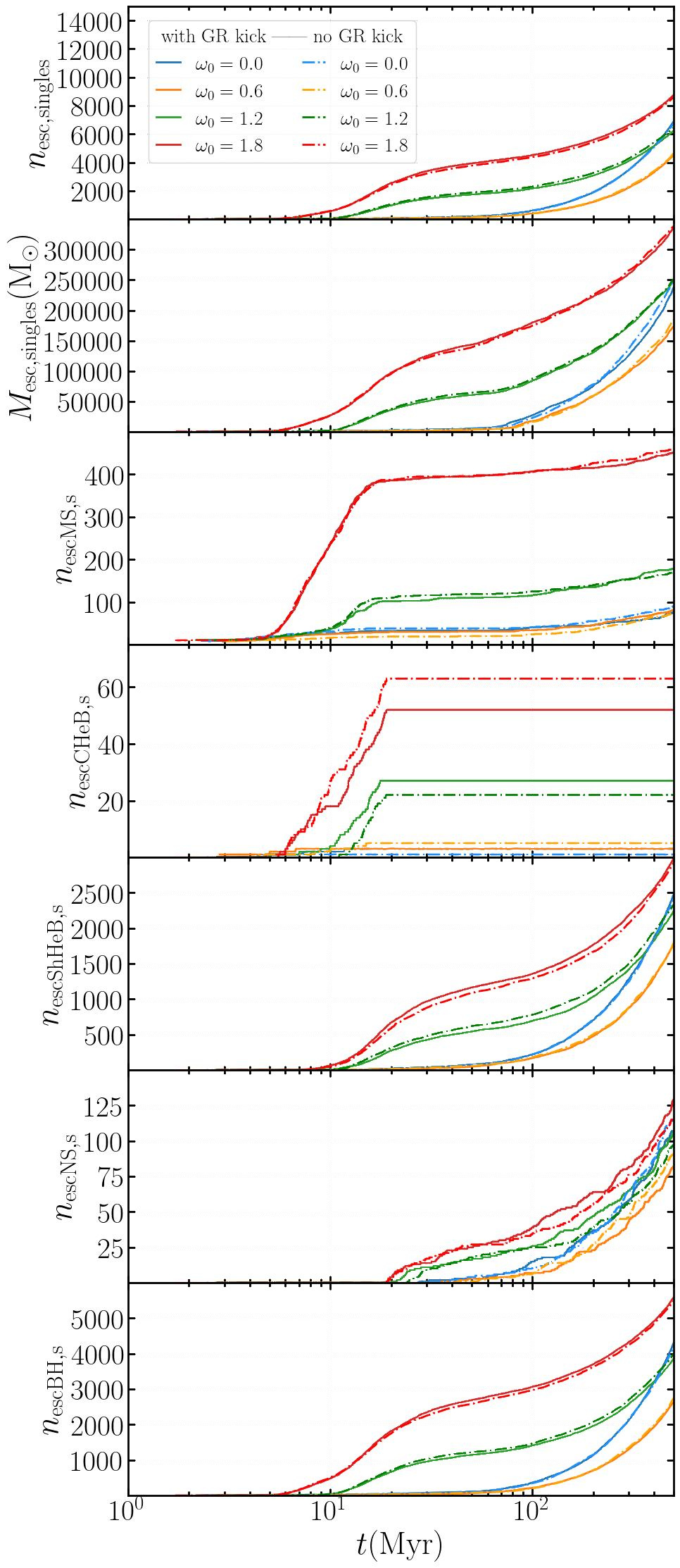}
    \caption{Escaping single stars in eight simulations: cumulative counts and masses. 
        Top panel: Cumulative number of escaping single stars, $n_{\mathrm{esc,singles}}$. 
        Second panel: Total mass in escaping single stars, $M_{\mathrm{esc,singles}}$. 
        Remaining panels: Cumulative counts of escaping single MS, CHeB, ShHeB, NS, and BH stars, $n_{\mathrm{escMS,s}}$, $n_{\mathrm{escCHeB,s}}$, $n_{\mathrm{escShHeB,s}}$, $n_{\mathrm{escNS,s}}$, and $n_{\mathrm{escBH,s}}$. 
        Solid curves show \textsc{K} models and dash-dotted curves show \textsc{NoK} models.
    }
    \label{escapers}
\end{figure}

Figure \ref{escapers} (single escapers), Fig. \ref{escapers_binaries} (binary escapers), and \tref{tab:escaper abundances} show that a higher initial bulk rotation, $\omega_{0}$, {is associated with earlier and stronger tidal mass loss in} both \textsc{NoK} and \textsc{K} models, most clearly for \textsc{(No)K}1.8 and \textsc{(No)K}1.2. The \textsc{(No)K}0.6 models {show a non-monotonic behaviour}, having $n_{\mathrm{esc,singles}}$, $n_{\mathrm{esc,binarymembers}}$ and $M_{\mathrm{esc,singles}}$ comparable to or below the non-rotating case. {A plausible physical interpretation is that two effects compete: (i) rotating King initial conditions are more compact than for their non-rotating counterparts (Sect. \ref{Initial_conditions}), which can reduce early tidal overflow, while (ii) stronger rotation accelerates early angular-momentum transport and structural readjustment (Sect. \ref{Section:Structural parameter evolution}), which enhances mass loss once the system expands and interacts with the tidal boundary. In this framework, $\omega_0=0.6$ can fall in an intermediate regime where the increased initial compactness dominates the net mass loss, whereas for larger $\omega_0$ the rotation-driven early evolution dominates.} This tidal field mass loss modifies cluster timescales and links directly to the structural and angular-momentum evolution discussed in Sects.~\ref{Section:Structural parameter evolution} and \ref{Section:Angular momentum transport}.

We then focus on the escaper counts by stellar types. The third panel of Fig. \ref{escapers} shows single MS escapers given by $n_{\mathrm{escMS,s}}$. The \textsc{(No)K}1.8 runs eject hundreds more MSs than the \textsc{(No)K}1.2 runs. Many of these MSs will form BHs via lCCSNe in the field (Sect. \ref{Section:Stars and compact objects}), and thus cannot fuel in-cluster hierarchical mergers. Although strong initial rotation speeds up the early collapse and gravogyro contraction (Fig. \ref{rlagr}; Sect. \ref{Section:Global dynamical evolution}), the simultaneous\footnote{Escapers are counted after they pass twice the current tidal radius, so the actual escape process starts earlier than the figure suggests.} loss of BH progenitors can hinder subsequent merger chains.

We next considered evolved phases, starting with escaping single CHeB stars ($n_{\mathrm{escCHeB,s}}$; Fig. \ref{escapers} panel four). Their behaviour resembles that of escaping MSs, though differences between \textsc{(No)K}1.2 and \textsc{(No)K}1.8 are reduced. Many escaping MSs rapidly become CHeBs, and likewise many CHeBs shortly become ShHeBs (see Fig. \ref{single_abundances_portrait}). After formation, most will remain in the ShHeB phase for a comparatively long time.

Escaping ShHeBs ($n_{\mathrm{escShHeB,s}}$) follow the overall $\omega_0$ trend, but show a pronounced increase in Phase IV (binary-heating dominated; Sect. \ref{Section:Global dynamical evolution}), echoing the earlier rise seen in $n_{\mathrm{escMS,s}}$. \tref{tab:escaper abundances} indicates that the \textsc{(No)K}0.6 runs yield the fewest single-star escapers across almost all types. {For the rarest subtypes (e.g. $n_{\mathrm{escCHeB,s}}\lesssim\mathcal{O}(1\text{--}10)$), Poisson counting noise can be large and individual model-to-model differences should be interpreted cautiously.}

For compact object escapers, $n_{\mathrm{escNS,s}}$ shows no strong model-to-model differences, though \textsc{(No)K}0.6 consistently yields the fewest single NS escapers. Escaping single BHs, $n_{\mathrm{escBH,s}}$, follow a similar behaviour: \textsc{(No)K}1.8 produces the largest number at 500~Myr (\tref{tab:escaper abundances}), while \textsc{(No)K}0.0 exhibits the strongest late rise in BH losses (Fig. \ref{escapers}). Because our kicks are fallback-dependent (Appendix~\ref{Sect: General relativistic merger recoil kicks}), and given the flat IMF (8-300~\msun{}) plus absent wind mass loss for BH progenitors, most BHs acquire such large fallback fractions that their natal kick velocities are $\approx 0$. At 500~Myr, the \textsc{(No)K}1.8 runs show the largest losses in compact objects; the \textsc{(No)K}0.0 runs display accelerating late-time loss that may overtake them, while the \textsc{(No)K}0.6 runs lose the least mass and the fewest compact objects.

About escaping binaries, Fig. \ref{escapers_binaries} and \tref{tab:binary_abundances} show that $n_{\mathrm{esc,binarymembers}}$ depends on initial rotation. The \textsc{(No)K}1.2 and \textsc{(No)K}1.8 models eject binary members earlier and, by 500~Myr, have roughly twice as many escaped binary members as the \textsc{(No)K}0.0 and \textsc{(No)K}0.6 runs. The latter two remain similar and do not display the large long-term differences seen for single escapers (Fig. \ref{escapers}).

For escaping binary components excluding compact objects, a few MS escapers (counted by $n_{\mathrm{escMS,b}}$) occur only in the \textsc{(No)K}1.2 and \textsc{(No)K}1.8 runs; they are mostly from primordial binaries, low-mass, and unlikely to form IMBHs. Binary CHeB escapers ($n_{\mathrm{escBH,b}}$) are rarer still.
Binary ShHeB losses ($n_{\mathrm{escShHeB,b}}$) are roughly twice as large and occur earlier in the $\omega_0=1.2,1.8$ models than in the $\omega_0=0.0,0.6$ models.
Many late-time escaping BHBH binaries contain two IMBHs, and thus cannot fuel hierarchical in-cluster growth. {The stronger-rotation models are therefore consistent with a reduced retained BH progenitor and seed population at late times.}

Overall, enabling GRKs produces the charted differences: with GRKs enabled the \textsc{(No)K}1.8 models show more CHeB single-star losses and fewer MS binary losses, while the \textsc{(No)K}1.2 models show the opposite trend. We again note that this is limited by small-number statistics.

\section{Summary and conclusions}
\label{Section:Summary and conclusions}

We present a set of a few pilot direct \nbody{} simulations that probe how bulk rotation and ultra-low metallicity with a top-heavy IMF shape Pop~III cluster dynamics. Rotation follows generalized rotating King models \citep{EinselSpurzem1999} with rigid core rotation and differentially decreasing near the half-mass radius and into the halo. \citet{Kamlahetal2022-rotation} analysed the general evolution and angular momentum transport in such models of Pop II star clusters; we find generally analogous dynamics, such as the gravogyro and gravothermal collapse, enhanced by the top-heavy IMF \citep[$8.0$--$300.0$~\msun{},][]{LazarBromm2022}. We study the impact of switching on and off GRKs (\citealt{dragon2-1}), and how the newly implemented stellar evolution for extremely metal-poor (EMP) stars \citep[here used $Z/\zsun{}=10^{-8}$,][]{Tanikawaetal2020} affects the formation and evolution of compact objects. 
While the mass function of Pop III star clusters is highly uncertain \citep[see Sect. \ref{Initial_conditions} and][]{Stacyetal2016,Liuetal2021a}, our pilot simulations model very massive Pop III star clusters (\tref{Initial_conditions}) to cover an unexplored parameter range and also because of stronger and faster dynamical effects, such as massive object formation, at high density \citep{Vergara2025a}. Direct \nbody{} simulations using the \petar{} code \citep{Wangetal2020d} for Pop III star clusters \citep{Liu2024,liu2024merging,liu2025on-the-formation} have used smaller cluster masses and particle numbers, requiring many independent realizations to average $N$ fluctuations. On the contrary, we used very large particle numbers ($1.01\times 10^5$ initially), which suffice to probe global dynamical effects of rotation and gravogyro-gravothermal collapse, but still provide a relatively small database for rare binaries and escapers (e.g. those involving BHs).
We ran the models for 500~Myr, using one non-rotating King model and three rotating ones (\citealt{EinselSpurzem1999}, $\omega_0=0.0,\,0.6,\,1.2,\,1.8$), each with and without GR merger recoil kicks.

{Primordial binary statistics are another major uncertainty for Pop~III clusters. In this work we fixed the initial binary fraction to $f_{\mathrm{b}}=1\%$ (\tref{Initial_conditions}), which should be regarded as a low-$f_{\mathrm{b}}$ limiting case rather than a best-estimate value. This choice suppresses IMBH formation pathways tied to primordial binary evolution (in particular Type-I and Type-III; Sect. \ref{Section:Stars and compact objects}) and, by construction, increases the relative importance of dynamical formation channels in our event budget. If $f_{\mathrm{b}}$ were higher, we would expect (i) a larger contribution of mergers from primordial binary channel, and (ii) stronger binary heating, which can delay gravothermal collapse and modify the balance between in-cluster hierarchical growth and dynamical ejections. Predicting the net change in the total IMBH production efficiency is therefore not straightforward without additional simulations, but the dominance of dynamical channels found here should be interpreted as an upper bound. A dedicated scan in $f_{\mathrm{b}}$ is left for future work.}

The global dynamical evolution can be divided into four phases (Figs.~\ref{Global_properties}, \ref{rlagr}, and \ref{avmass_l^2}):
\begin{itemize}
    \item Phase I (0-2~Myr): Global mass (\mcl{}) and tidal radius (\rtide{}) remain essentially constant because mass loss is negligible; however, initial bulk rotation ($\omega_0$) controls the structural response. Rapid angular-momentum transport occurs earlier and more strongly for larger $\omega_0$, producing a brief, synchronous bump in the Lagrangian radii near $\sim$0.1~Myr, inner-radius contraction and outer-radius expansion, and temporary rises then falls in \hmr{}, \rcore{}, and \mcore{} in the faster-rotating models. Low and zero rotation models show little change in \hmr{} over this phase.
    Also, $M_{\mathrm{group}}$ evolves similarly across mass groups, while $L_{\mathrm{group}}^2/L^2_{\omega_{0}=0.6,t=0}$ is larger for higher $\omega_0$, reflecting early outward angular-momentum migration.
    \item Phase II ($\sim$2-3~Myr): Rapid stellar-evolution mass loss (notably PISNe and compact-object formation) causes a sharp drop in total mass and \rtide{}, and produces a rapid increase in \hmr{} and \rcore{}. This is due to PISNe
    both remove mass and carry away a large fraction of the total angular momentum, while escapers (natal or GR kicks)
    and dynamical encounters add to mass loss and alter kinematics through fallback-dependent kicks. Measures of central segregation (\avmass{} within small \rlagr{}) show a similar qualitative behaviour across $\omega_0$ at this stage.
    \item Phase III ($\sim$3~Myr to $\sim$50-80~Myr; the latter is the core collapse time depending on $\omega_0$): Larger initial rotation shortens the gravothermal collapse timescale, leading to a higher central density and faster mass segregation. Post-phase II angular-momentum diagnostics (e.g. $L_{\mathrm{hCCSNe}}^2/L^2_{\omega_{0}=0.6,t=0}$) rise to a peak at $\sim$10-20~Myr in the \textsc{(No)K}1.2 and \textsc{(No)K}1.8 runs then decline. The \textsc{(No)K}1.8 models eject many stars.
    \item Phase IV (core collapse to 500~Myr, the simulations' end): Post-collapse evolution is driven by binary heating in the core, producing self-similar Lagrangian-radius expansion and overall cluster expansion; models with larger $\omega_0$ collapse earlier and expand more. The system becomes fully mass-segregated. Angular momentum is redistributed from high-mass to low-mass remnants. Low-mass BHs migrate to wider orbits. A compact central IMBH subsystem forms.
\end{itemize}

We find a {dependence of merger rates for stars and compact objects} on the initial rotation of Pop III clusters (Fig. \ref{time_evolution_ncoal_coll}), particularly visible in the \textsc{K} models. The largest initial rotation ($\omega_0=1.8$) corresponds to the {largest number of} coalescences and collisions (\tref{tab:coalescence_collision_rates}), including the most BHBH collisions. Concerning the BHBH collisions, large initial rotation does also seem to precipitate these, with the largest number of collisions found in \textsc{(No)K}1.8 simulations. 

All simulations produce (IM)BHs below, inside, and above the PISN mass gap (Figs.~\ref{BH_IFMR} and \ref{detailedIFMR}). The most massive IMBH has 943.19~\msun{} (from \textsc{NoK}1.2; Fig. \ref{BHBH_MSMS_MSBH_collisions}). Single-star core collapse can form IMBHs
above the gap; elsewhere IMBH formation requires binary channels — primarily MS-BH hyperbolic collisions, CHeB or ShHeB coalescence with BHs,
or MS-MS mergers in massive primordial binaries. Many BHs that cannot be produced by single star evolution instead form after a common-envelope phase and subsequent coalescence that produces a CHeB then ShHeB star, which then core-collapses into a BH. 

{With one realization per set-up, we separated our findings into (i) dynamically robust trends dominated by continuous quantities or large-number counts, including global structural evolution and timescales (core-collapse time and post-collapse self-similar expansion; Figs.~\ref{Global_properties} and \ref{rlagr}), angular-momentum transport signatures (Fig. \ref{avmass_l^2}), and total mass loss as well as total escaper numbers (\tref{tab:escaper abundances});
(ii) potentially stochastic trends dominated by small-number event statistics, such as specific collision or coalescence subtypes
and compact-object channels (e.g. BHBH collision counts; \tref{tab:coalescence_collision_rates}), rare escaper subtypes (e.g. $n_{\mathrm{escCHeB,s}}$), and single exceptional events (e.g. the lone BHNS merger). } 

Cluster evolution therefore reflects a competition: gravothermal-gravogyro contraction boosts early merger rates, while the loss of massive objects due to excess angular momentum suppresses subsequent IMBH growth and BH mergers. {Other physical effects, which may increase (IM)BH retention even for rotating clusters are:}
(i) {The presence of a} dark matter mini-halo around Pop III clusters, as considered by \citet[we note that those simulations did not include rotation]{Wangetal2022b}. {However, Pop III star clusters may also form in larger dark matter halos, and would be dominated by their baryons after cooling; see discussion in \cite{Kimmetal2016}.}
(ii) {Gas infall could add more mass to the central regions,} driving stronger contraction than predicted by gravothermal-gravogyro evolution. Such a contraction may trigger collapse of the central rotating IMBH subsystem and produce very massive seed BHs \citep[see][]{Kroupa2020}.

{Our results establish a general understanding of how initial rotation, angular momentum transport, and numerous massive objects in the top-heavy IMF affect global evolution, counts of compact objects, and their merger rates. In the future, we aim to \begin{enumerate*}[label=(\roman*)]\item continue the simulations for longer time, \item increase the number of simulations to improve the statistics of results, \item study the influence of an external potential (such as a dark matter mini-halo) on retention rates, and \item test a fractal initial structure, which may be present in the early phases \end{enumerate*}.}

\begin{acknowledgements}
    We thank the anonymous referee for their constructive comments that helped to improve the quality of this paper. 
    We thank Shuai Liu for helpful discussions and comments, and acknowledge computing time through the John von Neumann Institute for Computing (NIC) on the GCS Supercomputer JUWELS Booster.
        KW and RS acknowledge support by the German Science Foundation (DFG, project Sp 345/24-1).
        AT appreciates support by Grants-in-Aid for Scientific Research (17H06360, 19K03907, 24K07040, 25K01035) from the Japan Society for the Promotion of Science. 
        This material is based upon work supported by Tamkeen under the NYU Abu Dhabi Research Institute grant CASS.
        MCV acknowledges funding through ANID (Doctorado acuerdo bilateral DAAD/62210038) and DAAD (funding program number 57600326), and by International Max Planck Research School for Astronomy and Cosmic Physics at the University of Heidelberg (IMPRS-HD). 
        BL acknowledges the funding of the Deutsche Forschungsgemeinschaft (DFG, German Research Foundation) under Germany's Excellence Strategy EXC 2181/1 - 390900948 (the Heidelberg STRUCTURES Excellence Cluster).
        MAS acknowledges funding from the European Union’s Horizon 2020 research and innovation programme under the Marie Skłodowska-Curie grant agreement No. 101025436 (project GRACE-BH, PI Manuel Arca Sedda) and the support of the MERAC Foundation through the 2023 MERAC prize and support to research. 
        NN is grateful for funding by the Deutsche Forschungsgemeinschaft (DFG, German Research Foundation) -- Project-ID 138713538 -- SFB 881 (``The Milky Way System'', subproject B08).
    RS has been supported by NAOC/CAS Beijing, International Cooperation Office (2023-2025), by National Natural Science Foundation of China (NSFC) under grant No. 12473017, and by the CAS President's International Fellowship Initiative for Visiting Scientists (PIFI, 2026PVA0089).
    
\end{acknowledgements}

\bibliography{ref}
\bibliographystyle{aa}

\begin{appendix}

\section{Additional figures}

\begin{figure}
    \centering
        \includegraphics[width=0.9\columnwidth]{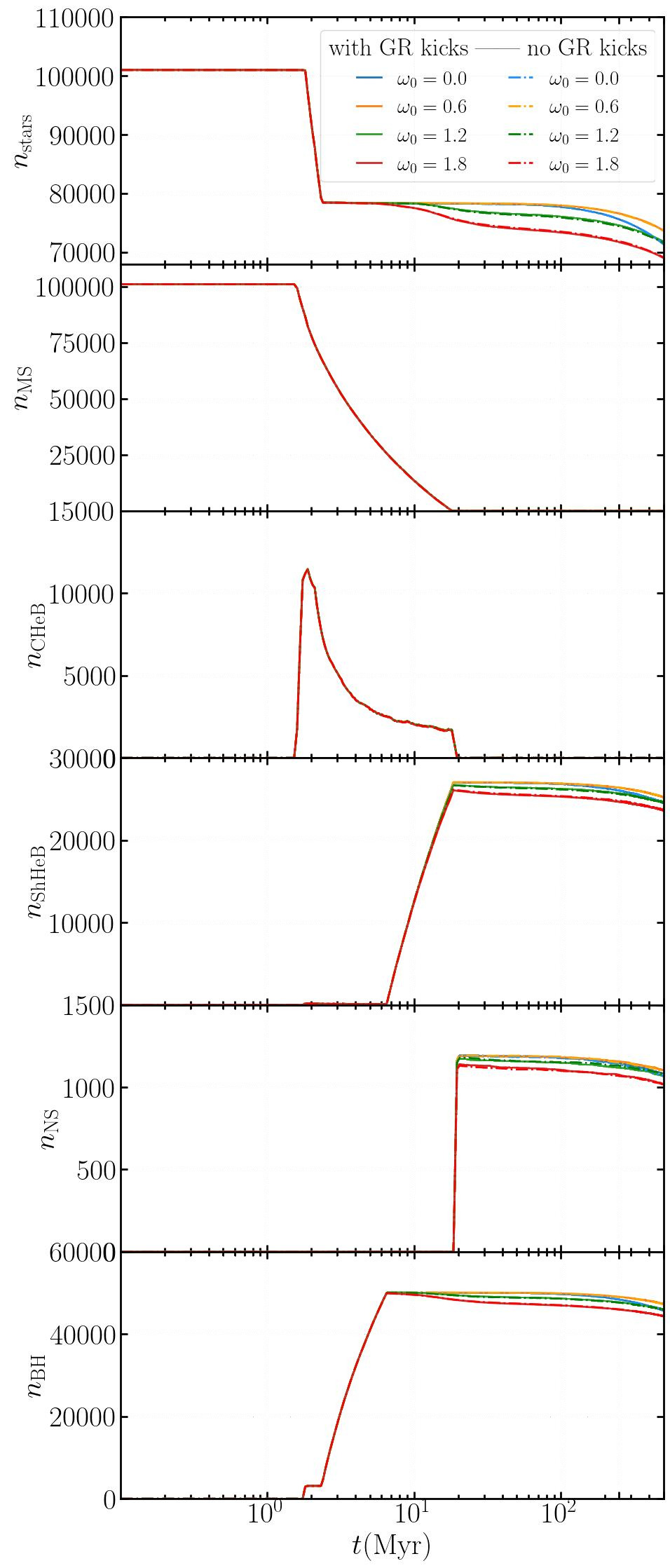}
        \caption{Six vertically stacked panels display the evolution of total star count and the counts of MS (main sequence star), CHeB (core-helium burning star), ShHeB (shell-helium burning star), NS (neutron star), and BH (black hole), for eight simulations. 
    }
        \label{single_abundances_portrait}
\end{figure}

\begin{figure}
    \centering
    \includegraphics[width=0.89\columnwidth]{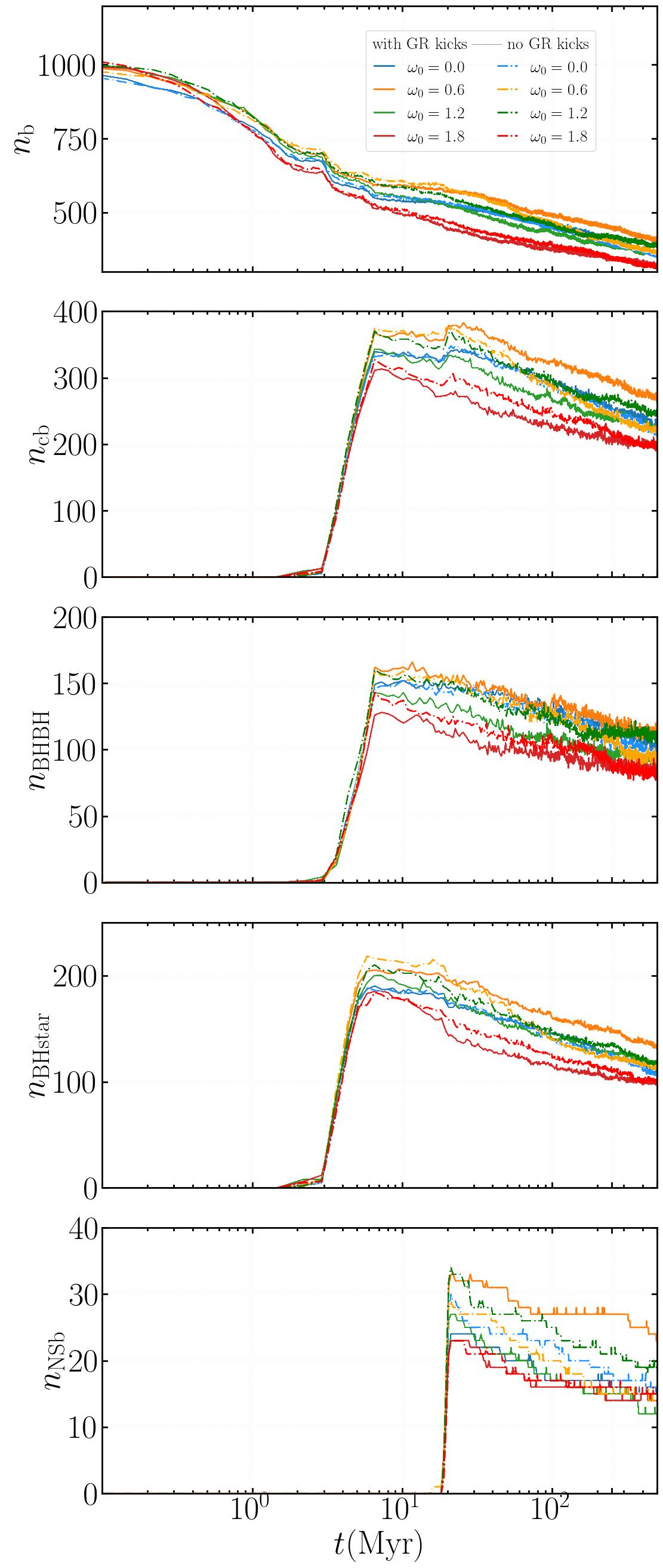}
    \caption{Top to bottom: Number of total binaries ($n_{\mathrm{b}}$), compact binaries ($n_{\mathrm{cb}}$), BHBHs ($n_{\mathrm{BHBH}}$), BH stars ($n_{\mathrm{BHstar}}$), and NS binaries ($n_{\mathrm{NSb}}$) for eight simulations.}
    \label{binary_abundances_portrait}
\end{figure}

\begin{figure*}
    \centering
        \includegraphics[width=0.9\textwidth]{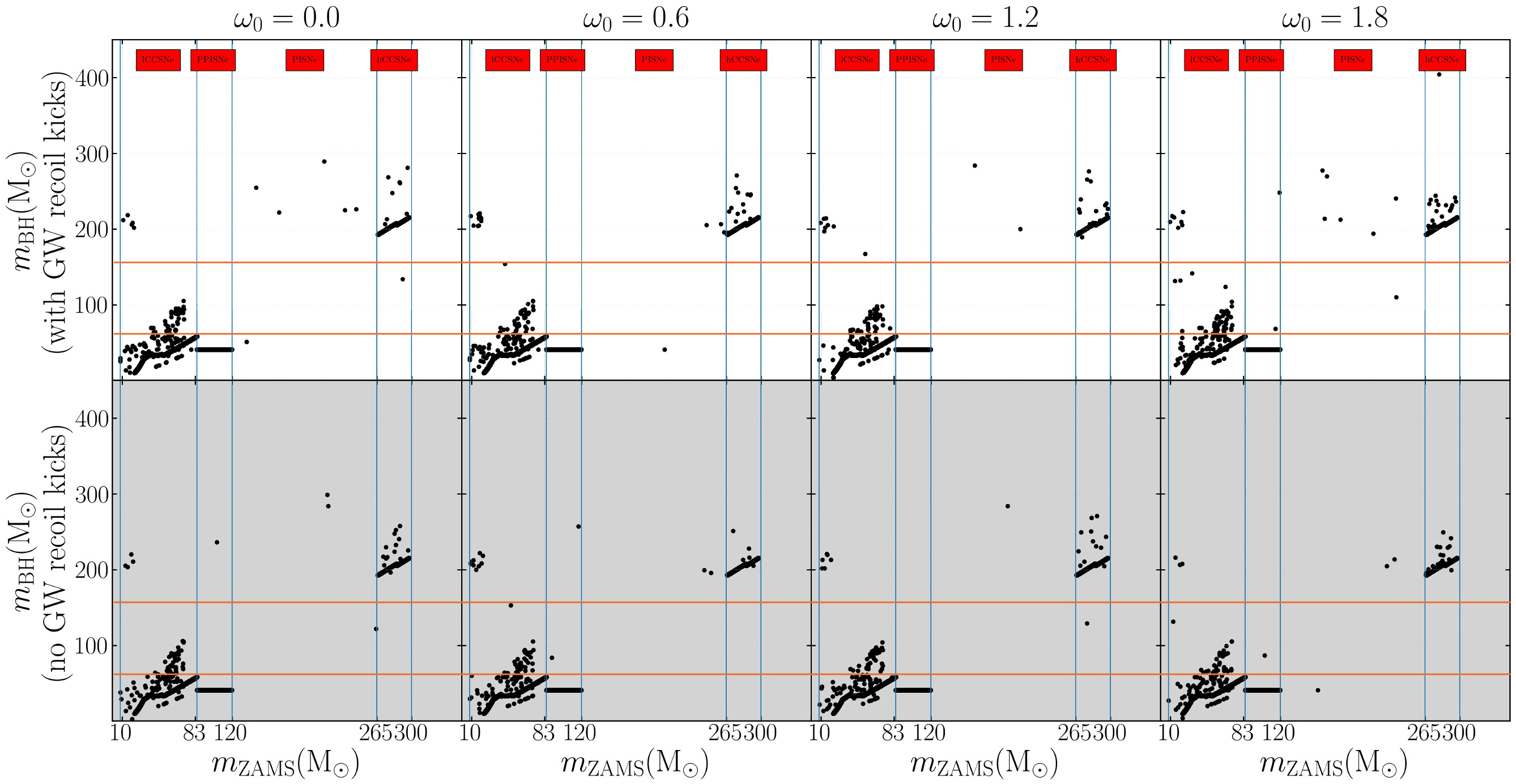}
        \caption{Initial-final-mass relation (IFMR): the BH mass $m_{\mathrm{BH}}$ versus the progenitor ZAMS star mass $m_{\mathrm{ZAMS}}$ for all eight simulations. Columns from left to right show increasing rotation $\omega_{0}=0.0,\,0.6,\,1.2,\,1.8$. \textsc{K} runs use a white background; \textsc{NoK} runs are shaded light grey. \textcolor{NavyBlue}{Blue} lines separate the progenitor-mass regimes, as annotated in \textcolor{red}{red}. Orange lines enclose the pair-instability mass gap ($\sim$64-161~\msun{}, \citealt{WoosleyHeger2021}). IMBHs appear below, within and above the PISN gap.}
        \label{BH_IFMR}
\end{figure*}

\begin{figure*}
        \includegraphics[width=\textwidth]{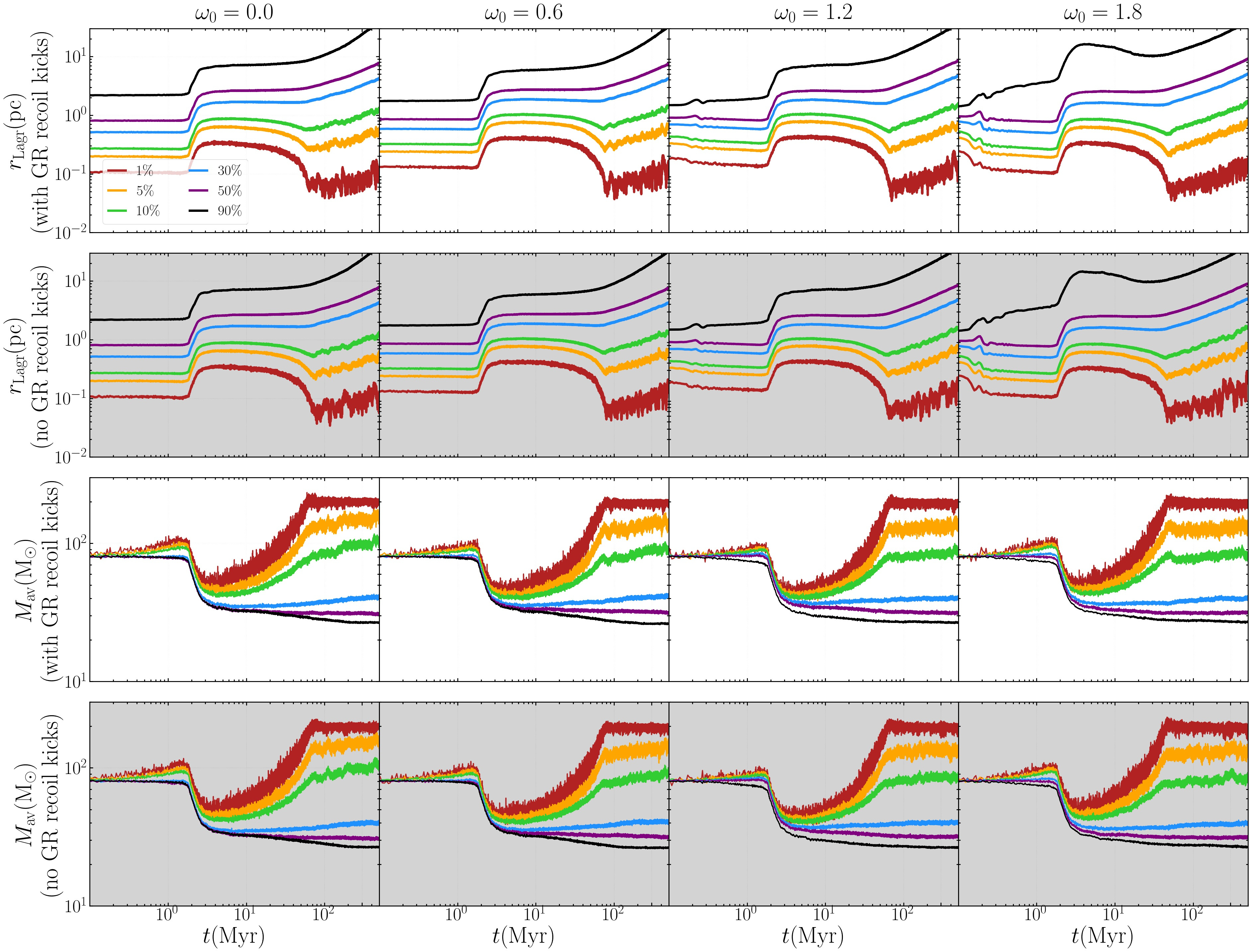}
        \caption{Lagrange radii (\rlagr{}) and average mass (\avmass{}) within spheres that contain 1\%, 5\%, 10\%, 30\%, 50\%, and 90\% of the total cluster mass at the current simulation time step for up to 500~Myr. Time is shown on a logarithmic scale to highlight the cluster's rapid early evolution. Each column represents one rotational parameter $\omega_{0}$ of the rotating King model ($\omega_{0}$ = 0.0, 0.6, 1.2, 1.8). Models with \textsc{K} are plotted on a white background; models without GR kicks (\textsc{NoK}) are shaded light grey.}
        \label{rlagr}
\end{figure*}

\begin{figure*}
    \includegraphics[width=\textwidth]{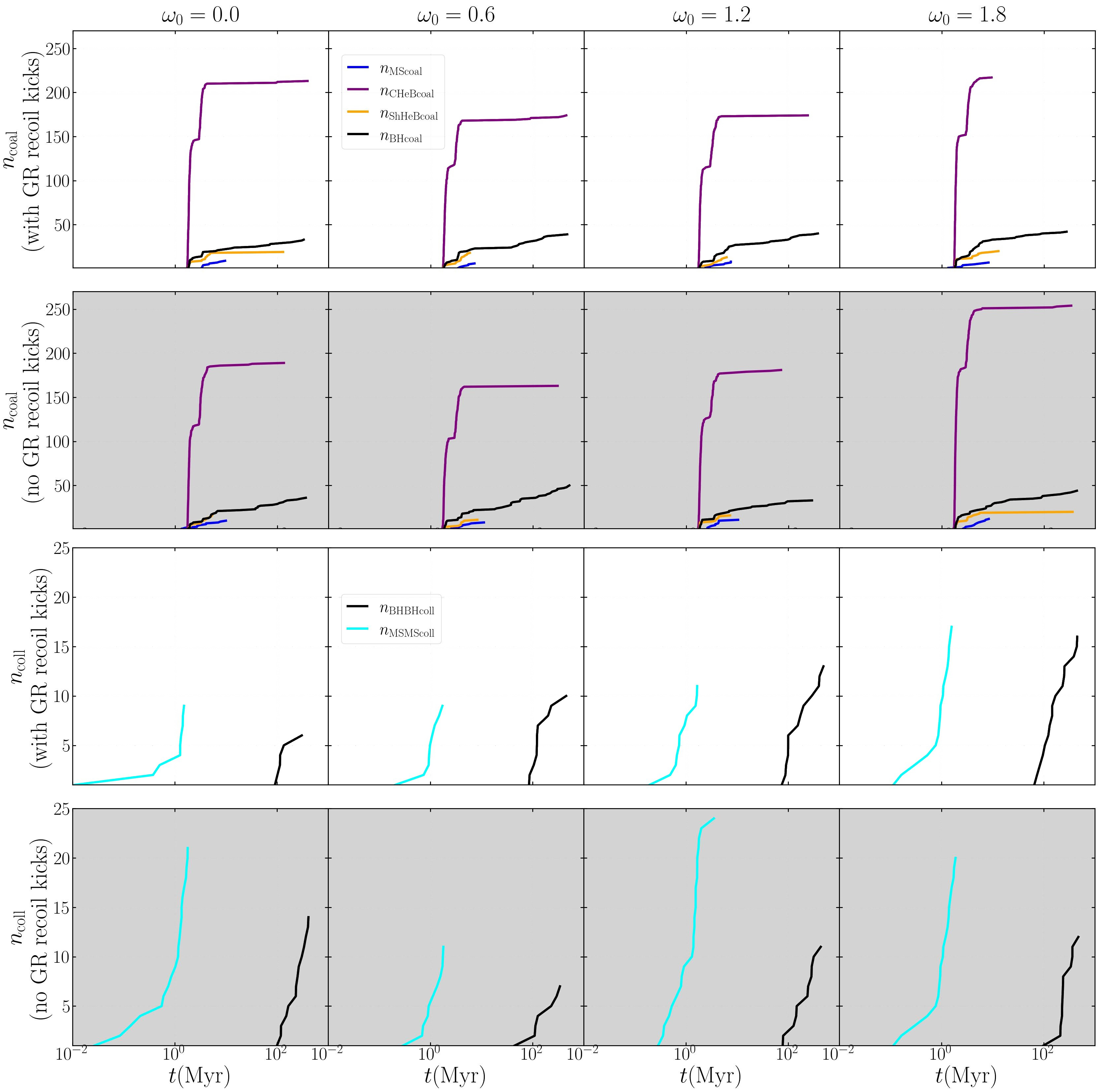}
    \caption{Time evolution of cumulative number of coalescences $n_{\mathrm{coal}}$ (top two rows) and collisions $n_{\mathrm{coll}}$ (bottom two rows) for the eight runs. Columns give increasing $\omega_{0}$ from left to right. \textsc{K} models (GR kicks on) use a white background; \textsc{NoK} models are shaded light grey. The horizontal axis is logarithmic in time (Myr) to clarify the pre-core-collapse phase. Colours indicate outcomes: MS (blue), CHeB (purple), ShHeB (orange), BH (black); collisions BH-BH (black) and MS-MS (cyan). Counts are in \tref{tab:coalescence_collision_rates}}
    \label{time_evolution_ncoal_coll}
\end{figure*}

\begin{figure}
    \centering
    \includegraphics[width=\columnwidth]{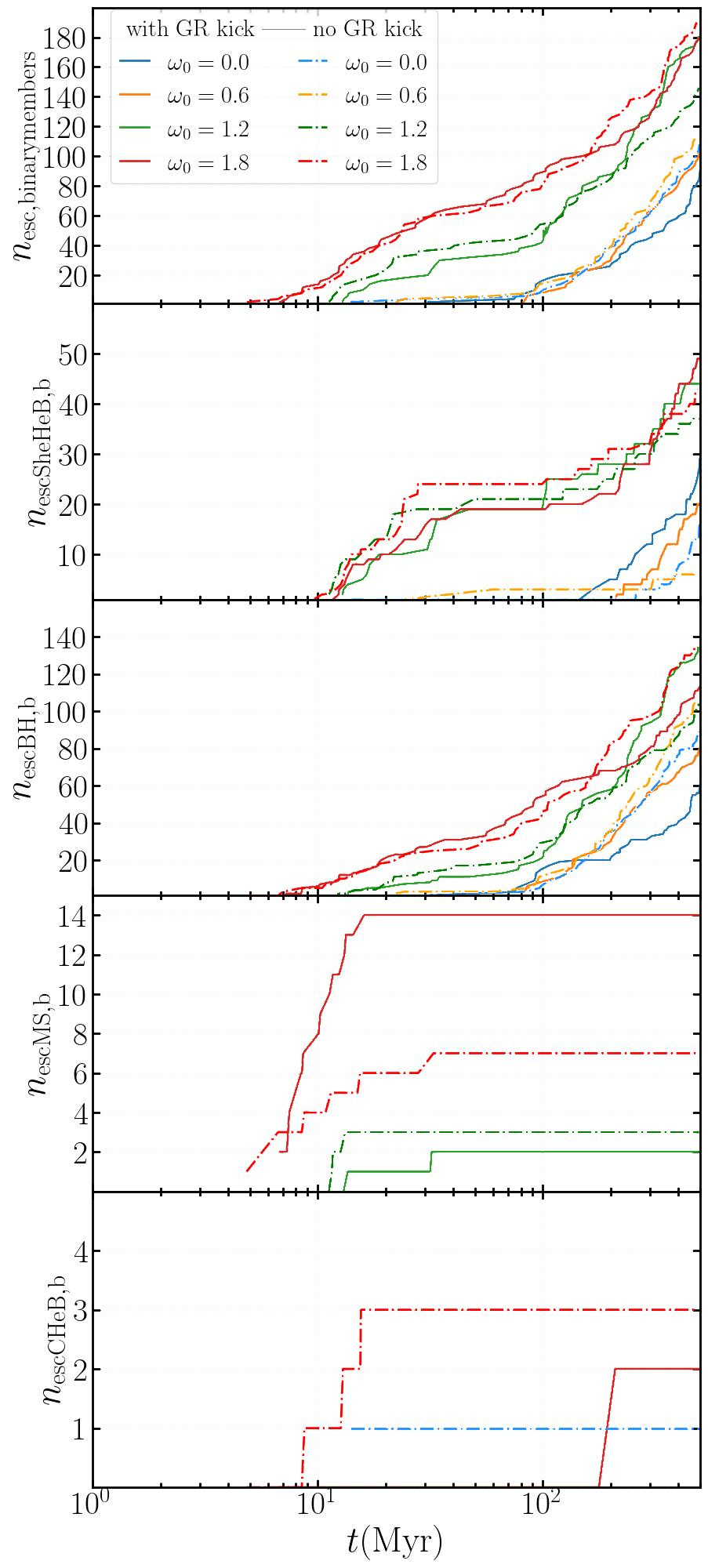}
    \caption{Cumulative counts of escaping binary members $n_{\mathrm{esc,binarymembers}}$ and of binary components: $n_{\mathrm{escMS,b}}$, $n_{\mathrm{escCHeB,b}}$, $n_{\mathrm{escShHeB,b}}$, and $n_{\mathrm{escBH,b}}$. \textsc{K}: solid curves; \textsc{NoK}: dash-dotted curves. 
    }
    \label{escapers_binaries}
\end{figure}    

\begin{figure*}
        \includegraphics[width=\textwidth]{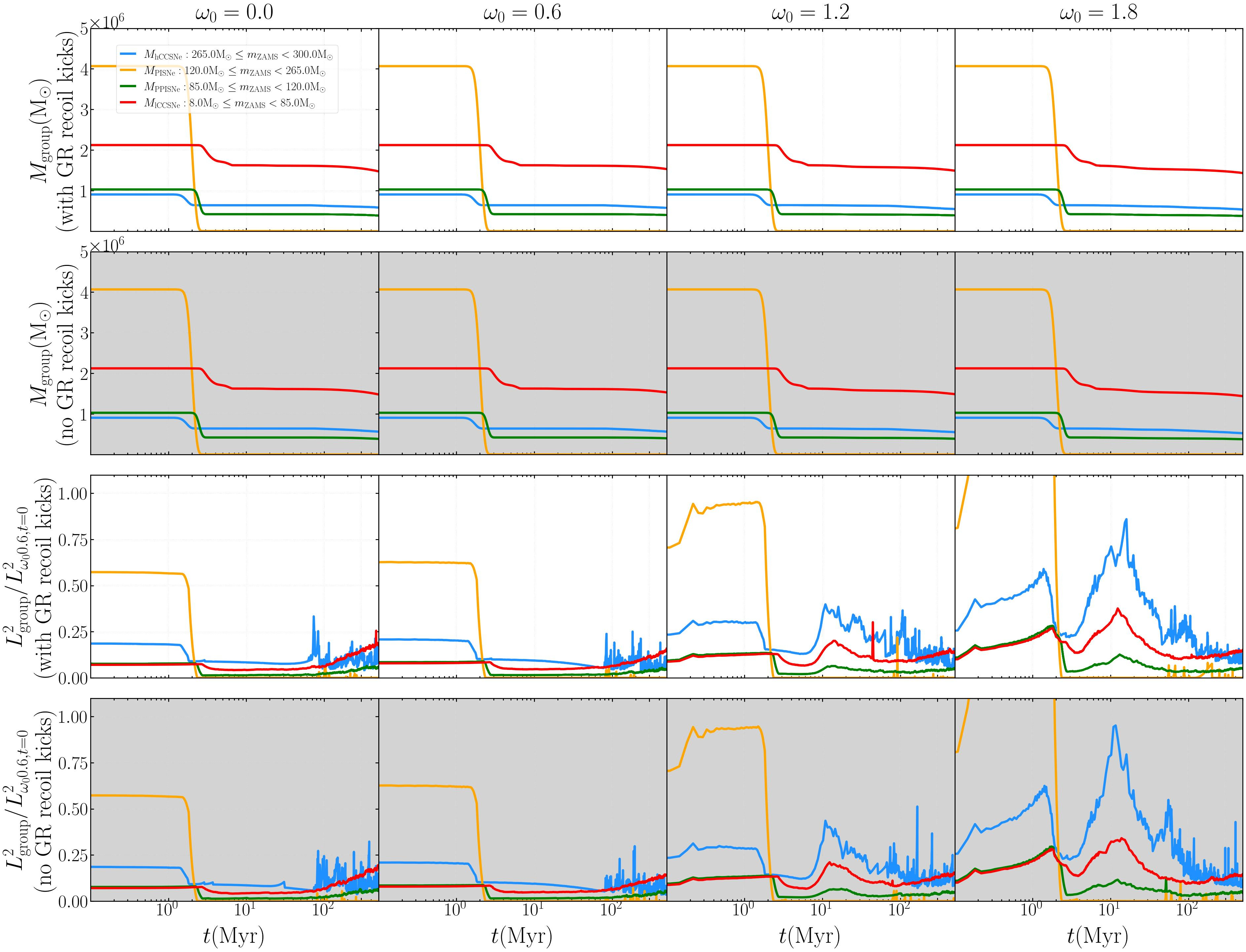}
    \caption{Top two rows: Total mass of the four mass groups ($M_{\mathrm{lCCSNe}}$ in \textcolor{red}{red}, $M_{\mathrm{PPISNe}}$ in \textcolor{ForestGreen}{green}, $M_{\mathrm{PISNe}}$ in \textcolor{orange}{orange}, $M_{\mathrm{hCCSNe}}$ in \textcolor{NavyBlue}{blue}). Bottom two rows: Each group's angular-momentum squared, $L_{\mathrm{group}}^2$, normalized by $L^2_{\omega_{0}=0.6,t=0}$ (i.e. $L_{\mathrm{group}}^2/L^2_{\omega_{0}=0.6,t=0}$). Columns show increasing $\omega_{0}$ from left to right ($\omega_{0}=0.0,\,0.6,\,1.2,\,1.8$). \textsc{K} runs are on a white background; \textsc{NoK} runs are shaded light grey.}
        \label{avmass_l^2}
\end{figure*}

\begin{figure*}
    \centering
    \includegraphics[width=0.84\textwidth]{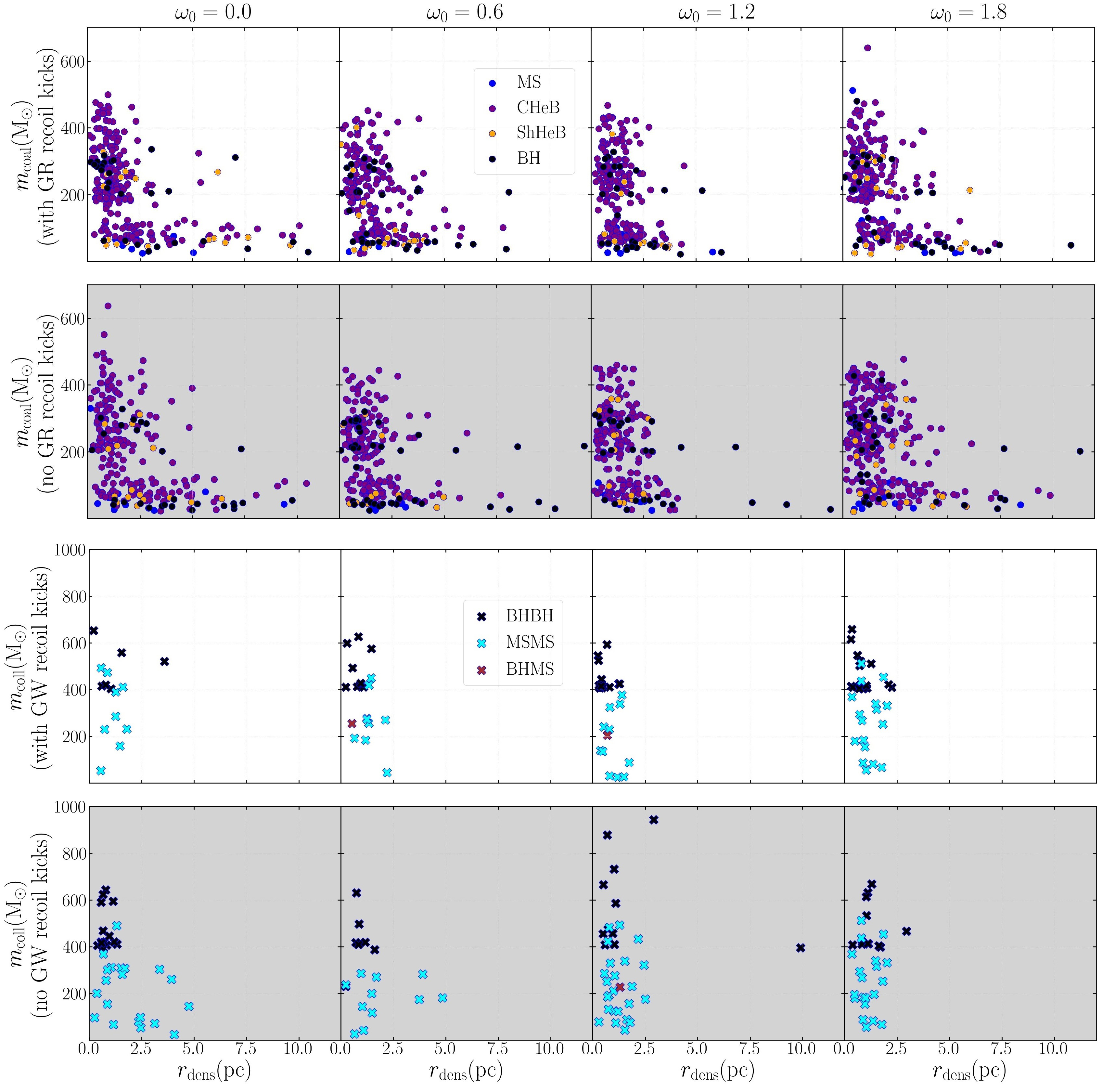}
    \caption{All coalescence (top two rows) and all collision events (bottom two rows) from the eight \nbody{} simulations. Plotted are product masses ($m_{\mathrm{coal}}$ or $m_{\mathrm{coll}}$) versus distance to the cluster density centre $r_{\mathrm{dens}}$. Columns from left to right show increasing rotation $\omega_{0}=0.0,\,0.6,\,1.2,\,1.8$. \textsc{K} models use a white background; \textsc{NoK} models are shaded light grey. Coalescence outcomes: MS (\textcolor{blue}{blue}), CHeB (\textcolor{purple}{purple}), ShHeB (\textcolor{orange}{orange}), BH (black). Collision markers: BH-BH (black cross), MS-MS (\textcolor{cyan}{cyan} cross), BH-MS (\textcolor{RedViolet}{violet} cross). Counts in \tref{tab:coalescence_collision_rates}.}
    \label{MS_GB_CHeB_ShHeB_NS_BH_mcoalescence_RI}
\end{figure*}

\begin{figure*}
    \centering
        \includegraphics[width=\textwidth]{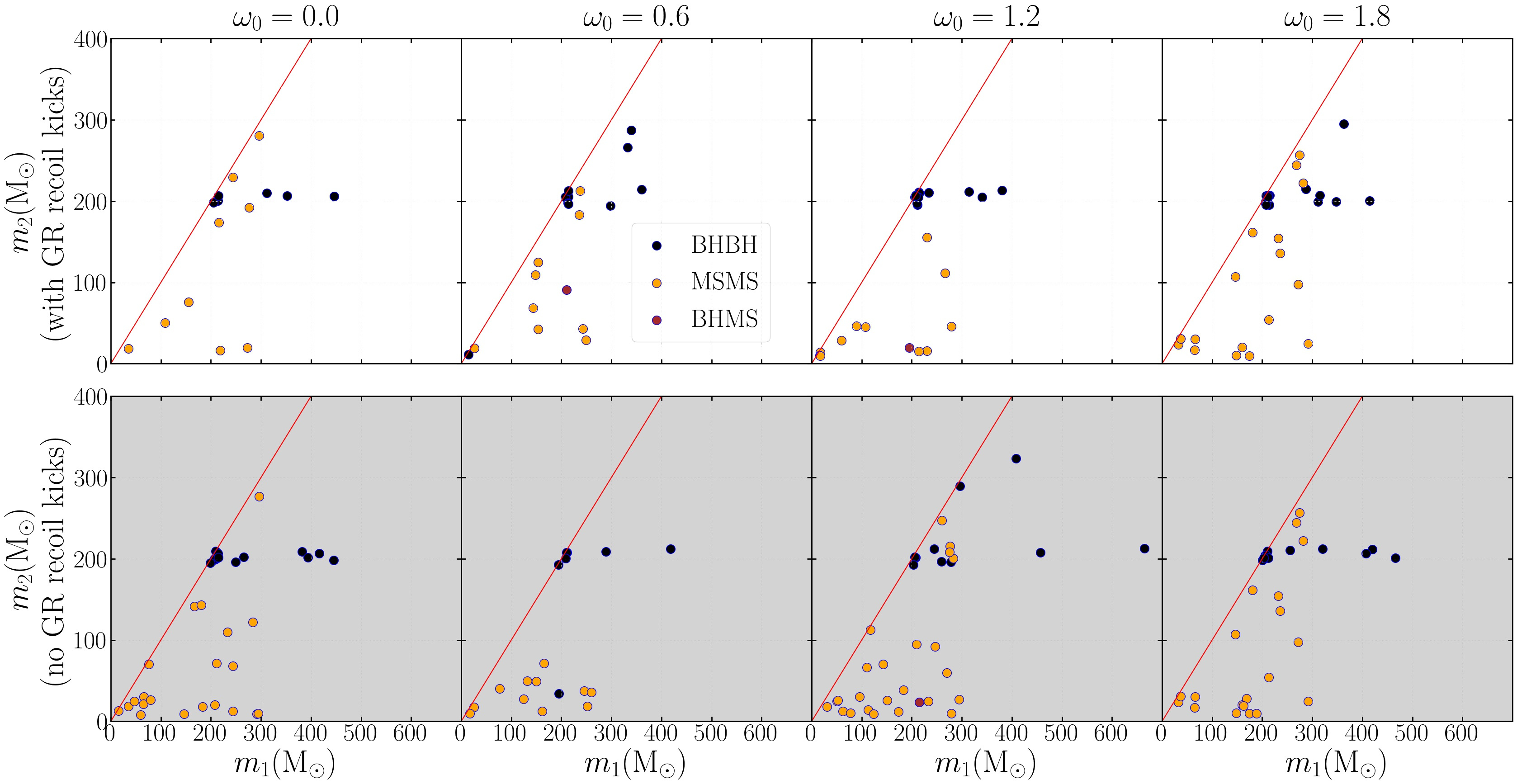}
        \caption{Primary vs\ secondary mass ($m_1>m_2$) for collisions from all eight simulations: MS--MS (\textcolor{orange}{orange}), BH--MS (\textcolor{purple}{purple}), BH--BH (\textcolor{black}{black}). Columns show $\omega_{0}=0.0,\,0.6,\,1.2,\,1.8$ left-to-right. \textsc{K} models are shaded white; \textsc{NoK} models are shaded grey. The \textcolor{red}{red} line denotes equal mass ratio ($q=1$). Counts in \tref{tab:coalescence_collision_rates}.}
        \label{BHBH_MSMS_MSBH_collisions}
\end{figure*}

\begin{figure}
    \centering
    \includegraphics[width=0.9\columnwidth]{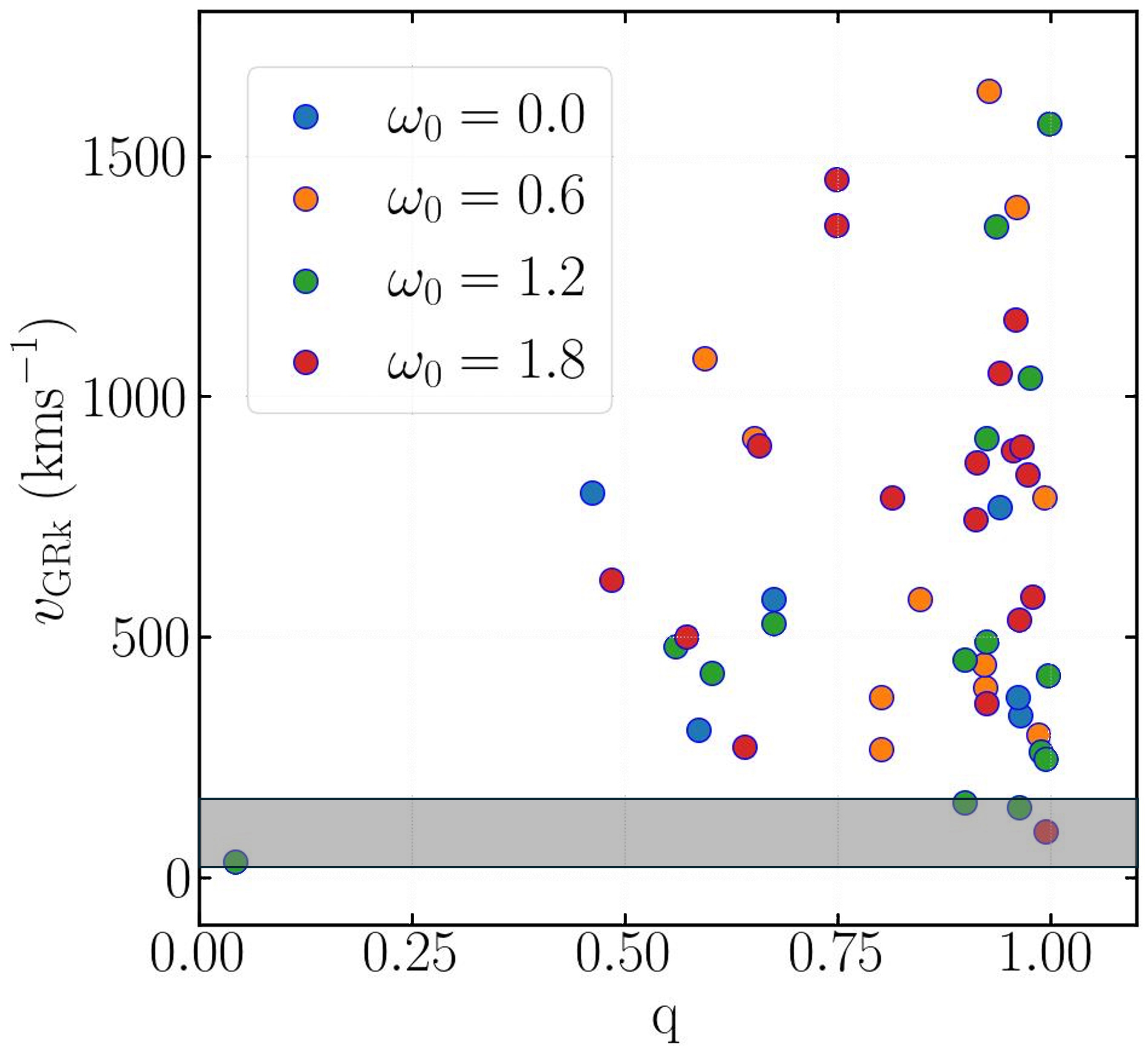}
    \caption{GRK velocity, $v_{\mathrm{GRk}}$, versus mass ratio, $q$, of the participating compact objects. All but one event are BHBH mergers; the exception is a BHNS merger, marked by an extreme $q$ and low $v_{\mathrm{GRk}}$. {The shaded part of the plot corresponds to the interval 26.9 -- 159.4 $\mathrm{kms}^{-1}$, the escape speeds range in cluster cores among all models.}}
    \label{q_vgw}
\end{figure}

\section{Tables of object counts}
\label{sec.table}

{All tables in this section report raw integer counts from a single realization per set-up. For any event or object-count $N$, the intrinsic counting uncertainty is approximately Poisson, with a reference 68\% ($\approx1\sigma$) error $\sigma_N\simeq\sqrt{N}$ (relative error $\simeq 1/\sqrt{N}$). For small $N$, confidence intervals are non-Gaussian and can be asymmetric; therefore, differences of a few events between models should be interpreted cautiously when $N$ is in the single-digit or few-tens regime.}

\begin{table*}
    \centering
        \caption{Remaining binary populations at 500~Myr.}
        \label{tab:binary_abundances}
        \begin{tabular}{|c|c|c|c|c||c|c|c|c|}
            \hline \textbf{Model ID} & \textsc{K}0.0 & \textsc{K}0.6 & \textsc{K}1.2 & \textsc{K}1.8 & \textsc{NoK}0.0 & \textsc{NoK}0.6 & \textsc{NoK}1.2 & \textsc{NoK}1.8 \\
            \hline \hline $n_{\mathrm{b}}$ & $355\pm19$ & $356\pm19$ & $390\pm20$ & $313\pm18$ & $358\pm19$ & $400\pm20$ & $350\pm19$ & $315\pm18$ \\
            \hline $n_{\mathrm{cb}}$ & $220\pm15$ & $212\pm15$ & $250\pm16$ & $195\pm14$ & $228\pm15$ & $263\pm16$ & $214\pm15$ & $192\pm14$ \\
            \hline $n_{\mathrm{BHBH}}$ & $98\pm10$ & $87\pm9$ & $113\pm11$ & $79\pm9$ & $104\pm10$ & $112\pm11$ & $85\pm9$ & $80\pm9$ \\
            \hline $n_{\mathrm{BHstar}}$ & $106\pm10$ & $111\pm11$ & $118\pm11$ & $101\pm10$ & $109\pm10$ & $129\pm11$ & $117\pm11$ & $98\pm10$ \\
            \hline $n_{\mathrm{NSb}}$ & $16\pm4$ & $14\pm4$ & $19\pm4$ & $15\pm4$ & $15\pm4$ & $22\pm5$ & $12\pm3$ & $14\pm4$ \\
            \hline
        \end{tabular}
        \tablefoot{Columns list the eight \nbo{} simulations. Rows give the numbers of remaining binaries $n_{\mathrm{b}}$, compact binaries $n_{\mathrm{cb}}$, BHBH binaries $n_{\mathrm{BHBH}}$, binaries containing a BH and a non-compact star $n_{\mathrm{BHstar}}$, and NS binaries $n_{\mathrm{NSb}}$. Notes on the uncertainty are given in Appendix~\ref{sec.table}.}
\end{table*}

\begin{table*}
    \centering
    \caption{Counts of coalescences and collisions in the eight simulations.}
    \label{tab:coalescence_collision_rates}
    \begin{tabular}{|c|c|c|c|c||c|c|c|c|}
        \hline \textbf{Model ID} & \textsc{K}0.0 & \textsc{K}0.6 & \textsc{K}1.2 & \textsc{K}1.8 & \textsc{NoK}0.0 & \textsc{NoK}0.6 & \textsc{NoK}1.2 & \textsc{NoK}1.8 \\
        \hline \hline $n_{\mathrm{MScoal}}$ & $9\pm3$ & $6\pm2$ & $8\pm3$ & $7\pm3$ & $10\pm3$ & $8\pm3$ & $11\pm3$ & $12\pm3$ \\
        \hline $n_{\mathrm{CHeBcoal}}$ & $213\pm15$ & $174\pm13$ & $174\pm13$ & $217\pm15$ & $189\pm14$ & $163\pm13$ & $183\pm14$ & $254\pm16$ \\
        \hline $n_{\mathrm{ShHeBcoal}}$ & $19\pm4$ & $18\pm4$ & $13\pm4$ & $20\pm4$ & $16\pm4$ & $11\pm3$ & $16\pm4$ & $20\pm4$ \\
        \hline $n_{\mathrm{BHcoal}}$ & $33\pm6$ & $39\pm6$ & $40\pm6$ & $42\pm6$ & $36\pm6$ & $50\pm7$ & $33\pm6$ & $44\pm7$ \\
        \hline $n_{\mathrm{totcoal}}$ & $274\pm17$ & $237\pm15$ & $235\pm15$ & $284\pm17$ & $251\pm16$ & $232\pm15$ & $243\pm16$ & $330\pm18$ \\
        \hline
        \hline $n_{\mathrm{MSMScoll}}$ & $9\pm3$ & $9\pm3$ & $11\pm3$ & $17\pm4$ & $21\pm5$ & $11\pm3$ & $24\pm5$ & $25\pm5$ \\
        \hline $n_{\mathrm{MSBHcoll}}$ & 0 & $1\pm1$ & $1\pm1$ & 0 & 0 & 0 & $1\pm1$ & 0 \\
        \hline $n_{\mathrm{BHBHcoll}}$ & $6\pm2$ & $10\pm3$ & $13\pm4$ & $16\pm4$ & $14\pm4$ & $7\pm3$ & $11\pm3$ & $12\pm3$ \\
        \hline $n_{\mathrm{totcoll}}$ & $15\pm4$ & $20\pm4$ & $25\pm5$ & $33\pm6$ & $35\pm6$ & $19\pm4$ & $35\pm6$ & $37\pm6$ \\
        \hline
        \hline $n_{\mathrm{totevents}}$ & $289\pm17$ & $257\pm16$ & $260\pm16$ & $319\pm18$ & $270\pm16$ & $267\pm16$ & $278\pm17$ & $367\pm19$ \\
        \hline
    \end{tabular}
    \tablefoot{The upper block lists coalescence products. The lower block lists collision progenitor pairs. Totals are $n_{\mathrm{totcoal}}$ for coalescences, $n_{\mathrm{totcoll}}$ for collisions, and $n_{\mathrm{totevents}}$ for their sum. Notes on the uncertainty are given in Appendix~\ref{sec.table}.} 
\end{table*}    

\begin{table*}[p]
    \centering
                \caption{Escaping-object counts at 500~Myr in the eight simulations.}
        \label{tab:escaper abundances}
        \begin{tabular}{|c|c|c|c|c||c|c|c|c|}
            \hline \textbf{Model ID} & \textsc{K}0.0 & \textsc{K}0.6 & \textsc{K}1.2 & \textsc{K}1.8 & \textsc{NoK}0.0 & \textsc{NoK}0.6 & \textsc{NoK}1.2 & \textsc{NoK}1.8 \\
            \hline \hline $n_{\mathrm{esc,singles}}$ & $6903\pm83$ & $4491\pm67$ & $6334\pm80$ & $8935\pm95$ & $6796\pm82$ & $4917\pm70$ & $6689\pm82$ & $8777\pm94$ \\
            \hline $M_{\mathrm{esc,singles}}$(\msun{}) & 238416 & 173845 & 252120 & 342161 & 252089 & 192843 & 258546 & 343595 \\
            \hline $n_{\mathrm{escMS,s}}$ & $79\pm9$ & $82\pm9$ & $181\pm13$ & $454\pm21$ & $91\pm10$ & $74\pm9$ & $171\pm13$ & $459\pm21$ \\
            \hline $n_{\mathrm{escCHeB,s}}$ & $4\pm2$ & $4\pm2$ & $28\pm5$ & $53\pm7$ & $2\pm1$ & $6\pm2$ & $22\pm5$ & $64\pm8$ \\
            \hline $n_{\mathrm{escShHeB,s}}$ & $2482\pm50$ & $1775\pm42$ & $2283\pm48$ & $3046\pm55$ & $2448\pm49$ & $1912\pm44$ & $2431\pm49$ & $2987\pm55$ \\
            \hline $n_{\mathrm{escNS,s}}$ & $107\pm10$ & $84\pm9$ & $109\pm10$ & $133\pm12$ & $116\pm11$ & $96\pm10$ & $103\pm10$ & $119\pm11$ \\
            \hline $n_{\mathrm{escBH,s}}$ & $4310\pm66$ & $2628\pm51$ & $3914\pm63$ & $5703\pm76$ & $4230\pm65$ & $2903\pm54$ & $4133\pm64$ & $5607\pm75$ \\
            \hline 
            \hline $n_{\mathrm{esc,binarymembers}}$ & $90\pm9$ & $102\pm10$ & $180\pm13$ & $180\pm13$ & $108\pm10$ & $112\pm11$ & $146\pm12$ & $190\pm14$ \\
            \hline $n_{\mathrm{escMS,b}}$ & 0 & 0 & $2\pm1$ & $14\pm4$ & 0 & 0 & $3\pm2$ & $7\pm3$ \\
            \hline $n_{\mathrm{escCHeB,b}}$ & 0 & 0 & 0 & $2\pm1$ & $1\pm1$ & 0 & 0 & $3\pm2$ \\
            \hline $n_{\mathrm{escShHeB,b}}$ & $29\pm5$ & $20\pm4$ & $44\pm7$ & $49\pm7$ & $16\pm4$ & $6\pm2$ & $37\pm6$ & $43\pm7$ \\
            \hline $n_{\mathrm{escNS,b}}$ & $1\pm1$ & $2\pm1$ & 0 & $2\pm1$ & $2\pm1$ & $1\pm1$ & $2\pm1$ & $2\pm1$ \\
            \hline $n_{\mathrm{escBH,b}}$ & $60\pm8$ & $80\pm9$ & $134\pm12$ & $113\pm11$ & $89\pm9$ & $105\pm10$ & $104\pm10$ & $135\pm12$ \\
            \hline
        \end{tabular}
                \tablefoot{The upper block gives escaping single stars: total $n_{\mathrm{esc,singles}}$, MS $n_{\mathrm{escMS,s}}$, CHeB $n_{\mathrm{escCHeB,s}}$, ShHeB $n_{\mathrm{escShHeB,s}}$, NS $n_{\mathrm{escNS,s}}$, and BH $n_{\mathrm{escBH,s}}$; the corresponding escaping single-star mass is listed as $M_{\mathrm{esc,singles}}$. The lower block gives escaping binary members: total $n_{\mathrm{esc,binarymembers}}$, MS $n_{\mathrm{escMS,b}}$, CHeB $n_{\mathrm{escCHeB,b}}$, ShHeB $n_{\mathrm{escShHeB,b}}$, NS $n_{\mathrm{escNS,b}}$, and BH $n_{\mathrm{escBH,b}}$. Notes on the uncertainty are given in Appendix \ref{sec.table}.}
\end{table*}

\section{Methodology details}
\subsection{\nbo{}}
\label{sec:nbo}

\nbo{}\footnote{\url{https://github.com/nbody6ppgpu}} is optimized for large-scale computing clusters by using MPI \citep{Spurzem1999}, SIMD, OpenMP and GPU \citep{NitadoriAarseth2012,Wangetal2015} parallelization techniques. In combination with the Kustaanheimo-Stiefel (KS) regularization \citep{Stiefel1965} for binary stars and close encounters, the Hermite scheme with hierarchical block time steps \citep{McMillan1986,Hutetal1995,Makino1991a,Makino1999} and the Ahmad-Cohen (AC) neighbour scheme \citep{AhmadCohen1973}, two-body and chain regularization \citep{MikkolaTanikawa1999a,MikkolaTanikawa1999b,MikkolaAarseth1998} the code allows for highly accurate simulations to be made of star clusters of a realistic size, including binaries.

The AC scheme divide the gravitational forces acting on it into the regular component, originating from distant stars, and an irregular part, originating from nearby stars ("neighbours"). Regular forces, efficiently accelerated on the GPU, are updated in larger regular time steps, while neighbour forces are much more fluctuating and need updates in shorter time intervals. Since neighbour numbers are usually small compared to the total particle number, their implementation on the CPU using OpenMP \citep{Wangetal2015} provides the best overall performance. Post-Newtonian dynamics of relativistic binaries is currently still using the orbit-averaged Peters \& Matthews formalism \citep{PetersMathews1963,Peters1964}, as described e.g. in  \citet{DiCarloetal2019,DiCarloetal2020a,DiCarloetal2020b,DiCarloetal2021,Rizzutoetal2021a,Rizzutoetal2022,ArcaSeddaetal2021a}, but many upgrades were made in \citet{dragon2-1} and this work.

We refer interested readers to \citet{spurzem2023rev} for a detailed review of \nbo{}.

\subsection{\mcluster{} and \fopax{}: Initialization of rotating star clusters}
\label{Section:McLuster,fopax}

\begin{enumerate}
    \item The star clusters models are first initialized with \mcluster{} \citep{Kuepperetal2011a,Kamlahetal2022-preparing,Levequeetal2022a}. 
        The \mcluster{} output models are ready as an input for \nbo{} as initial models. The input parameters are given in Sect. \ref{Section:Initial conditions} and \tref{Initial_conditions}. 

    \item Then, we generate a 2-D Fokker-Planck initial models using \fopax{} \citep{EinselSpurzem1999,Kimetal2002,Kimetal2004,Kimetal2008}. 
        The code produces a 2-D mesh based output of density $\rho$ and velocity dispersions $\sigma$ as a function of $r$ and $z$ (as cylindrical coordinates) based on the rotating King model $f(E, J_{\mathrm{z}})$ that are characterized by a pair of parameters $(W_{0},\omega_{0})$:
        $f_{\mathrm{rk}}~\propto~\left(\mathrm{e}^{\beta E}-1\right)\times~\mathrm{e}^{-\beta\Omega_{0}J_{\mathrm{z}}},$
        where $\beta = 1/(m\sigma_{\mathrm{c}}^2)$ and the dimensionless rotation parameter is given by $\omega_{0}=\sqrt{9/4\times \pi Gn_{\mathrm{c}}}\times \Omega_{0}$. $\sigma_{\mathrm{c}}$, $n_{\mathrm{c}}$ are central one-dimensional velocity dispersion and the central density, respectively. Potential-density pairs \citep[e.g.][]{Binney2008} for these models are created by relating $\beta$ to the King parameter $W_0$ via $W_0 = \beta m(\psi-\psi_{\mathrm{t}})$, where $\psi$ and $\psi_{\mathrm{t}}$ are the central King potential and the King potential at the truncation radius, which is normally characterized by \rtide{}, the tidal radius \citep[for numerical and computational methods of this step, see also][]{Henyeyetal1959,Cohn1979,Spurzem1994,Spurzem1996}.
    
    \item We use a Monte Carlo rejection technique to generate a discrete system of $N$ particles, following the known distributions of $\rho$ and $\sigma$. The output, ready for \nbo{}, consist of mass, the 3-D position, velocity data of each particle. This creates a cluster with a differential rotation based on the rejection technique (it includes different rotating shells, which separate a certain number of particles, based on the used King model, $\omega_0$ and the number of particles). 
    Finally, all data is scaled to standard H\'enon units. As a result, we have an initial star cluster model with rotating King model \nbody{} distribution, the chosen IMF, and relevant binary orbital parameter distributions conserved from \mcluster{}.
\end{enumerate}

\subsection{\textsc{SSE, BSE,} and Pop III stellar evolution}
\label{Sect: SSE, BSE, and Pop III stellar evolution}

The role of stellar mass is extremely important, especially at larger masses.
For stars less massive than 8~\msun{}, we use the classical stellar evolution recipes of \citet{Hurleyetal2000,Hurleyetal2002b} with recent upgrades in \citet{Kamlahetal2022-preparing}. Adjustments for these recipes are applied where extremely metal-poor (EMP; Pop III) stars evolve differently due to the absence of heavy elements. Therefore, we assume EMP stars have negligible mass loss from winds and pulsations \citep{Nakauchietal2020} at extremely low metallicities. Since we use top-heavy IMFs, thus using numerous stars with $M/\msun > 8$, the classical stellar evolution recipes fail to model Pop III stars.
In this paper, we present the implementation into \nbo{} of the extended fitting formulae derived from the fitting to 1-D \textsc{HOSHI} stellar evolution models \citep{Takahashietal2016,Takahashietal2018,Takahashietal2019,Yoshidaetal2019} of extremely metal-poor stars \citep{Tanikawaetal2020,Tanikawaetal2021a,Tanikawaetal2021c,Hijikawaetal2021}. At the metallicity that we adopt in this work, $Z/\zsun{}=10^{-8}$, the fitting formulae by \citet{Tanikawaetal2020} are generally valid from $8$ to $1280$~\msun{}. We adopt the extrapolation procedure of \citet{Wangetal2022b} for stars outside this range. Extrapolation to stellar mass beyond $1280$~\msun{} till several $10^3$~\msun{} can be done with these formulae, because no abrupt changes in the fitting are expected for these large masses (e.g. stellar luminosity is almost proportional to stellar mass and the star lifetime is quite similar at these masses) 
Extrapolation is more difficult below $8~\msun{}$, where stellar mass has significantly greater effects on stellar evolution. For stars with masses below $8~\msun{}$, we adopt the evolutionary prescriptions of \citet{Hurleyetal2000} at the lowest available metallicity of $Z/\zsun{}=10^{-2}$ 
Since only a small fraction of stars fall into this mass range under our initial conditions (see \tref{Initial_conditions}), this choice does not significantly affect the accuracy of our models. To summarize, besides \citet{Wangetal2022b}, we are among the first groups to combine direct \nbody{} simulation with full Pop III stellar evolution from \citet{Tanikawaetal2020}. 

We briefly highlight the implementations in the evolution of the stars for $Z/\zsun{}=10^{-8}$, imported from \citet{Tanikawaetal2020}. 
Normally, the stellar evolution is divided into five distinct phases in chronological order: Main Sequence (MS), Hertzsprung gap (HG), Core-Helium Burning (CHeB), Shell-Helium Burning (ShHeB) and the remnant phases of either NS or BH. In addition, \citet{Tanikawaetal2020} define the BSG and RSG phases that relate to the stellar surface, i.e. the effective temperature $T_{\mathrm{eff}}$, of the stars. Stars with $\mathrm{log}_{10}(T_{\mathrm{eff}}/K) \geq 3.65$ are classified as BSGs. Likewise, stars with $\mathrm{log}_{10}(T_{\mathrm{eff}}/K) < 3.65$ are classified with RSGs.

The BSG phase begins at ZAMS and ends when $\mathrm{log}_{10}(T_{\mathrm{eff}}/K) < 3.65$. Subsequently, the RSG phase begins, which ends at the time of carbon ignition, which is the end point of the 1-D simulations used for the fitting formulae. For example, depending on $T_{\mathrm{eff}}$ a CHeB star can become a BSG or a RSG. 
For metallicity of $Z/\zsun{}=10^{-8}$, as shown in Tab.~2 of \citealt{Tanikawaetal2020}:
\begin{itemize}
    \item Stars with $8 < M/\msun{} < 13$ have entered into their ShHeB phases by the time they become RSG stars.
    \item Stars with masses of $13 \leq M/\msun{} < 50$ end with BSG stars.
    \item Stars with $M/\msun{} \geq 50$ are still CHeB stars when they become RSG stars.
    \item In the case of binary stars, it is possible that the H envelope of the CHeB star is fully stripped by tides, Roche Lobe overflow (RLOF) or CEE changing the type of the star to a naked Helium MS (HeMS) star. For these stars, available fitting formulae of lowest metallicity of $Z/\zsun{}=5\times 10^{-3}$ from \citet{Hurleyetal2002b} are used instead.
\end{itemize}

{
We refer readers to \citet{Tanikawaetal2020} for detailed stellar evolution parameters, or check our code publicly available on GitHub\footnote{\url{https://github.com/nbody6ppgpu/Nbody6PPGPU-beijing/tree/EMP}}.
Using a time step parameter $\eta=0.01$ \citep{spurzem2023rev}, the local error scales as $\sqrt{\eta}$. The relative energy error is $\sim 10^{-8}$ per \nbody{} time unit ($\sim 10^4$~yr). Due to the uncorrelated nature of these errors, the global accumulation is slow, reaching only $\sim 10^{-6}$ after 100~Myr. Given the system's chaotic nature, this precision is sufficient. Moreover, iterative Hermite schemes can further improve energy conservation by achieving near time-symmetry \citep[e.g.][]{kokubo1998on-a-time-symmetric, Glaschkeetal2014, AmaroSeoaneetal2014, spurzem2023rev}.
}

\subsection{General relativistic merger recoil kicks (GRKs)}
\label{Sect: General relativistic merger recoil kicks}
The latest studies of IMBH growth \citep[e.g.][]{DiCarloetal2019,DiCarloetal2020a,DiCarloetal2020b,DiCarloetal2021,Rizzutoetal2021a,Rizzutoetal2022} and star cluster dynamics with \nbo{} \citep[e.g.][]{Kamlahetal2022-preparing,Kamlahetal2022-rotation} do not include a GR merger recoil treatment (in addition to missing PN terms). \citet{ArcaSeddaetal2021a} included GRKs via a posteriori analysis, but this does not replace a fully self-consistent treatment during the simulation. The updated \nbo{} presented in this paper and \textsc{Nbody7} \citep{Aarseth2012,Banerjeeetal2020,Banerjee2021a} contain a proper treatment of such kicks \citep[also see][]{dragon2-1}. These velocity kicks depend on spins and mass ratio, due to asymmetric GW radiation during the final inspiral and merger. Numerical relativity (NR) models \citep{Campanellietal2007,Rezollaetal2008,Hughes2009,vanMeteretal2010b,JimenezFortezaetal2017} have been used to formulate semi-analytic descriptions for \textsc{MOCCA} and \textsc{Nbody} code family \citep{Morawskietal2018,Morawskietal2019,Banerjee2021a,BelczynskiBanerjee2020,ArcaSeddaetal2021a,Banerjee2021c}. The smaller the spin (e.g. \textsc{Fuller} model, (nearly) non-spinning BHs), the larger the kick velocity. The larger the mass ratio, the smaller the kick velocity \citep{Morawskietal2018,Morawskietal2019}. 
In the other extreme case with non-aligned natal spins and small mass ratios, the asymmetry in the GW may produce GR merger recoils that reach of order $1000~\kmps{}$ \citep{Bakeretal2008,vanMeteretal2010b}. 

Typically, the BHBH orbital angular momentum sets the angular momentum budget for the post-merger BH, so the final spin tends to align with the orbital angular momentum \citep{Banerjee2021a}. By contrast, in physical collisions or mergers during binary-single interactions, the orbital angular momentum does not dominate, and therefore the final BH spin can be low. \citet{Banerjee2021a} also includes a treatment for random isotropic spin alignment of dynamically formed BHs. Additionally, \citet{Banerjee2021a} sets the GRK to zero for NSNS and BHNS mergers \citep{ArcaSedda2020,Chattopadhyayetal2021}, but applies numerical-relativity fitting formulae from \citet[updated by \citealt{Banerjee2021c}]{vanMeteretal2010b} to assign kicks to BHBH merger products. The final spin of the merger product is then evaluated in the same way as a BHBH merger. 

The GRK-enabled models of this work, for NS-NS, BH-BH, and NS-BH mergers, we draw the remnant spin from a Maxwellian distribution with $\sigma=0.2$ and apply a mass loss due to GW emission: the remnant mass is 98.5\% of the summed pre-merger compact object masses.

For further details of the implementation, we refer readers to \citet{ArcaSeddaetal2020a} and \citet{dragon2-1}.

\end{appendix}

\end{document}